\documentclass[pdflatex,sn-mathphys-num]{sn-jnl}

\usepackage{graphicx}%
\usepackage{multirow}%
\usepackage{amsmath,amssymb,amsfonts}%
\usepackage{amsthm}%
\usepackage{mathrsfs}%
\usepackage[title]{appendix}%
\usepackage{xcolor}%
\usepackage{textcomp}%
\usepackage{manyfoot}%
\usepackage{booktabs}%
\usepackage{algorithm}%
\usepackage{algorithmicx}%
\usepackage{algpseudocode}%
\usepackage{listings}%
\usepackage{soul}
\usepackage{enumitem}

\begin{document}

\title{Graph-Based Learning for Multi-Horizon Martian Atmospheric Forecasting}

\author*[1]{\fnm{Gary} \sur{Myler}}\email{garymyler@gmail.com}

\author[1]{\fnm{James} \sur{Holmes}}

\author[1]{\fnm{Manish} \sur{Patel}}

\author[1]{\fnm{Amel} \sur{Bennaceur}}

\affil*[1]{\orgname{The Open University}, \orgaddress{\country{UK}}}

\abstract{\textbf{Purpose.} Martian weather forecasting is important for future exploration, but atmospheric behaviour on Mars combines spatial, temporal, vertical, and dust-driven processes in ways that challenge current modelling and forecasting approaches. Existing machine learning studies often reduce this structure to local time series, which limits their ability to capture wider atmospheric dynamics.

\textbf{Methods.} This paper introduces a graph-based data engineering framework called MaGMA (Martian Graph-based Multi-horizon Atmospheric Forecasting) that transforms OpenMARS reanalysis fields into structured learning objects for Martian atmospheric forecasting. The framework represents local atmospheric patches as graph nodes and links them through neighbouring regions, successive time steps, longer temporal dependencies, and dynamically similar atmospheric states. It integrates recent atmospheric history, engineered physical descriptors, and vertical atmospheric information to support forecasting across multiple horizons.

\textbf{Results.} We evaluate  MaGMA across five unseen Martian years, including regular years and a global dust storm year. In regular years, the model achieves overall coefficients of determination of approximately 0.73--0.85, indicating that it captures a substantial proportion of the variation in the target atmospheric variables. For dust-column forecasting, the model outperforms classical and deep temporal baselines in most year--horizon comparisons. In the global dust storm year, dust-column prediction remains strong at shorter horizons, with coefficients of determination above 0.8 for the first two horizons, but broader multivariate performance declines, showing that extreme regimes still challenge generalisation.

\textbf{Conclusion.} The study shows that graph-based data engineering can create reusable and diagnostically useful representations for planetary atmospheric forecasting. It also identifies two priorities for future work: improving learning under rare extreme regimes and making better use of vertical atmospheric structure.}




\maketitle

\section{Introduction}
\label{sec:intro}
\begin{quote}
\emph{The purpose of computing is insight, not numbers.}\\
\hfill Richard W. Hamming
\end{quote}

Mars has a dynamic atmosphere in which dust, radiation, temperature, pressure, and wind interact across spatial and temporal scales~\cite{SanchezLavega2024MartianDynamics}. Atmospheric forecasting is important for both Mars science and exploration:  (i) it supports the study of climate variability, circulation, seasonal behaviour, and dust storm development~\cite{Holmes2020OpenMARS,Kahre2006DustCycle} and (ii) it  informs mission operations, including observation planning, energy management, rover traversability, and instrument use~\cite{Cantor2019Mars2020Dust}. As Mars missions generate increasingly rich atmospheric datasets, forecasting methods must exploit these data in ways that preserve their structure and support reuse across studies~\cite{Holmes2020OpenMARS,Wilkinson2016FAIR}.

Dust makes this problem particularly difficult~\cite{He2024MartianDustStorms,Kahre2006DustCycle}. Local and regional dust events affect radiative heating, circulation, surface temperature, and atmospheric temperature, often producing wider atmospheric changes~\cite{Gebhardt2021DustRadiationFeedback,Kahre2006DustCycle}. Global dust storms represent an extreme regime in which dust loading expands across much of the planet~\cite{He2024MartianDustStorms,Guzewich2019MSLGDS}. Their onset and evolution remain difficult to forecast because they emerge from nonlinear interactions among surface lifting, background circulation, radiative feedback, and evolving dust distributions~\cite{Newman2021DustStormPrediction,Gebhardt2021DustRadiationFeedback}. Distinguishing routine dust variability from global dust storm behaviour therefore remains a central challenge for Martian atmospheric modelling~\cite{He2024MartianDustStorms,Newman2021DustStormPrediction}.

This challenge is also a data engineering problem. Reanalysis and global circulation model datasets are not simple tabular datasets. They are multivariate, spatially gridded, vertically layered, and temporally evolving. Forecast-relevant signals may occur in the recent history of a location, in neighbouring regions, in the vertical atmospheric column, or in dynamically related regions elsewhere on the planet~\cite{Holmes2020OpenMARS,Yigit2023MartianWholeAtmosphere}. If a learning workflow flattens these fields into independent feature vectors or isolated time series, it may discard the relationships that carry the forecasting signal before model training even begins~\cite{Reichstein2019DeepLearningEarth,Yu2024SpatiotemporalEarthScience}.

Existing machine learning workflows for weather and climate prediction often rely on fixed feature sets, local time series, or gridded inputs that do not explicitly encode relationships among neighbouring regions, vertical layers, and dynamically similar atmospheric states~\cite{Schultz2021MLWeatherClimate,DeBurghDay2023MLWeatherClimate}. Earlier work on Martian weather prediction showed that machine learning models can forecast several OpenMARS dynamical variables from local time series, but that dust storm behaviour remains difficult when the problem is reduced to one-dimensional sequence forecasting. This limitation motivates a shift from local time-series prediction to structured spatiotemporal learning. For example, dust opacity at one location may depend on its own recent history, neighbouring wind conditions, vertical temperature gradients, and dust states elsewhere. The transformation from scientific data product to learning representation therefore becomes part of the forecasting method~\cite{Reichstein2019DeepLearningEarth,Wilkinson2016FAIR}.

This paper introduces \emph{MaGMA}, a Martian Graph-based framework for Multi-horizon Atmospheric forecasting. The MaGMA framework transforms multidimensional OpenMARS reanalysis fields into graph-structured learning objects that preserve spatial, temporal, vertical, and atmospheric-state relationships. Nodes represent local atmospheric patches over temporal windows. Edges encode spatial neighbourhoods, temporal continuity, longer temporal dependencies, and similarity between dynamically related atmospheric states. This representation allows the model to learn from both local atmospheric history and the wider relational structure of the Martian atmosphere~\cite{Kipf2017GCN,Schlichtkrull2018RGCN,Yu2024SpatiotemporalEarthScience}.

The MaGMA framework combines temporal sequence encoding, engineered atmospheric descriptors, vertical column information, relation-aware graph propagation, adaptive routing, and multi-horizon forecast heads. It predicts multiple atmospheric variables across short, medium, and longer forecast horizons. It also supports diagnostic analysis of the information pathways that drive prediction, including recent temporal history, engineered features, graph structure, and vertical atmospheric information~\cite{Holmes2020OpenMARS}.

The evaluation trains the model on one Martian year containing a global dust storm and tests it on complete unseen Martian years, including regular years and a separate global dust storm year~\cite{Fedorova2024MY28GDS,Guzewich2019MSLGDS}. This design assesses cross-year generalisation and tests whether the model captures dust-related atmospheric structure beyond common atmospheric regimes~\cite{Guzewich2019MSLGDS,Reichstein2019DeepLearningEarth}.

MaGMA advances Martian atmospheric forecasting from model selection over preprocessed time series to the design of reusable learning representations: it transforms OpenMARS data fields into structured graph objects and shows how this representation supports both multi-horizon prediction and diagnosis under regular and extreme atmospheric regimes.
More specifically, the contributions of the paper are as follows:

\begin{itemize}[leftmargin=*]
\item \textit{Graph-based data engineering for Martian reanalysis.}
We define a pipeline that transforms OpenMARS reanalysis fields into reusable spatiotemporal graph learning objects, preserving the spatial, temporal, vertical, and atmospheric-state structure needed for forecasting.

\item \textit{Multi-relation atmospheric graph representation.}
We construct a graph representation in which nodes encode local atmospheric patches over temporal windows, and edges capture spatial neighbourhoods, temporal continuity, longer-range temporal dependencies, and dynamically similar atmospheric states.

\item \textit{Multi-horizon graph neural forecasting model.}
We use a graph neural architecture to learn from the relational structure encoded in the atmospheric graph. The model integrates temporal history, engineered atmospheric descriptors, vertical column information, and relation-aware propagation to predict multiple atmospheric variables across short, medium, and longer forecast horizons.

\item \textit{Cross-year evaluation under regular and extreme regimes.}
We evaluate the framework across unseen Martian years, including a global dust storm stress test, and compare dust-column forecasting against linear, tree-based, recurrent, and temporal convolutional baselines.

\item \textit{Diagnostic analysis of forecasting information pathways.}
We analyse how temporal, engineered, vertical, and graph-based components contribute to performance, showing how the framework supports both prediction and interpretation of planetary atmospheric learning systems.
\end{itemize}

The remainder of the paper is organised as follows. Section~\ref{sec:relatedWork} reviews related work on Martian atmospheric modelling, machine learning for weather forecasting, and graph-based spatiotemporal learning. 
Section~\ref{sec:dataset} describes the OpenMARS dataset and the atmospheric variables used in this study. 
Section~\ref{sec:magma} introduces the MaGMA framework and its overall data engineering and forecasting workflow. 
Section~\ref{sec:graph} details the construction of the spatiotemporal graph, including node features, edge relations, and forecasting targets. Section~\ref{sec:gnn} presents the graph neural forecasting architecture. Section~\ref{sec:experimetns} describes the training setup, baselines, and evaluation protocol. 
Section~\ref{sec:results} reports and discusses the forecasting results, dust-storm stress test, and component sensitivity analysis. 
Finally, Section~\ref{sec:conclusion} concludes the paper and outlines directions for future work.

\section{Related Work}
\label{sec:relatedWork}
This section positions MaGMA with three different strands of related work, time-series forecasting, machine learning approaches to climate prediction and Martian atmospheric forecasting. Time-series methods provide the most direct baseline for the earlier OpenMARS forecasting setup and for the temporal only models in this study. However, the central gap addressed by MaGMA is representational, not just architectural: Mars Reanalysis fields are spatially gridded, vertically structured, temporally evolving and dynamically coupled. For this reason, the section also reviews recent graph-based and global data-driven weather models that preserve relational structure across atmospheric fields. The final part of the section discusses Martian weather and dust-storm forecasting, with particular attention to the limitations of single-site or one-dimensional forecasting workflows.

\paragraph{Weather forecasting}
Weather forecasting has a long history on Earth, where recent work has increasingly explored machine learning (ML) as a complement to numerical weather prediction. Early ML-based studies often focused on local or station-level prediction tasks. For example, existing work~\cite{earth_weather_1} uses a deep neural network for point-wise rain classification from seven variables, including humidity, temperature, pressure, and rain, collected at a local weather station in Japan. Each variable is processed through fully connected layers, concatenated into a shared representation, and passed through a softmax layer for binary rain classification. Although the forecast horizon is limited to one hour, the deep learning (DL) model outperforms traditional ML methods such as eXtreme Gradient Boosting (XGBoost) and support vector machines. Similarly, existing work~\cite{transductive_earth_weather_4} uses a transductive variant of long short-term memory (LSTM) to predict temperature in five European cities using data from Weather Underground\footnote{\url{https://www.wunderground.com/}}, and shows improved performance over vanilla LSTMs. Other work~\cite{hewage_earth_weather_2} uses local weather station data to show that temporal convolutional networks (TCNs) can outperform vanilla LSTM models.

More recent work has moved from local time-series forecasting towards global, structured, data-driven weather prediction using reanalysis datasets such as ERA5. GraphCast~\cite{lam2022graphcast}, developed by Google DeepMind, uses an encoder--decoder architecture based on Graph Neural Network (GNN) layers. Its encoder represents the Earth through a high-resolution multi-mesh graph, while its decoder maps the learned multi-mesh representation back to a latitude--longitude grid. GraphCast has been shown to outperform the European Centre for Medium-Range Weather Forecasts (ECMWF) High Resolution Forecast (HRES), a leading numerical weather prediction (NWP) system, for several surface and vertical atmospheric variables. FourCastNet~\cite{kurth2023fourcastnet} also targets global forecasting from ERA5, combining architectures such as Vision Transformers (ViT)~\cite{vision_transformer} and Fourier neural operators~\cite{li2021fourier} to model high-resolution atmospheric fields and learn spatial dynamics. Pangu-Weather~\cite{pangu-weather} similarly uses a three-dimensional architecture based on a variant of ViT to represent atmospheric input fields and reports lower Root Mean Square Error than both FourCastNet and the Integrated Forecasting System (IFS) NWP model.

Together, these studies show a shift from isolated station-level forecasting towards models that preserve spatial, temporal, and vertical structure in atmospheric data. This shift is directly relevant to Martian forecasting: OpenMARS is also a multidimensional reanalysis product, and reducing it to independent local time series risks discarding relationships that carry forecasting information. MaGMA builds on this direction by constructing graph-structured learning objects from OpenMARS fields, so that local atmospheric histories, neighbouring regions, vertical information, temporal links, and dynamically similar states can be represented together before forecasting.

\paragraph{Weather Forecasting for Mars}
Weather forecasting for Mars has a shorter history than terrestrial weather forecasting, and it remains strongly shaped by the available observing systems, reanalysis products, and physical models. Observations from landers and orbiters provide important constraints on Martian atmospheric behaviour, but they are sparse compared with Earth observations and vary across instruments, spatial coverage, and temporal sampling. The InSight lander is one such example: it collected local measurements of surface pressure, near-surface air temperature, and dust optical depth during a regional dust storm, showing how lander observations can support the interpretation of larger-scale atmospheric behaviour and future predictive capability~\cite{banfield2020atmosphere_of_mars_nature}.

Dust-storm prediction remains one of the central challenges for Martian forecasting. Recent reviews of dust-storm forecasting approaches~\cite{montabone2018forecasting_dust_storm_mars, forget2017atmospheric_dust_mars_review} argue that methods based mainly on domain knowledge and statistical analysis are not yet sufficient for accurate and timely forecasting. A recurring recommendation is to make better use of spacecraft observations through data assimilation, where observations are incorporated into atmospheric models to improve estimates of the atmospheric state. This has already been shown to improve forecasts for specific atmospheric quantities, such as carbon monoxide~\cite{holmes19}. These studies highlight both the importance of observational constraints and the difficulty of forecasting dust-driven atmospheric behaviour from limited and heterogeneous data.

Machine learning has only recently been explored for Martian weather prediction. Ishaani and Puri~\cite{priya21} used maximum temperature observations from NASA's Curiosity rover over 5.5 Earth years and compared the prediction errors of several ML models. Their study demonstrated the potential of data-driven methods for Martian weather variables, but it remained a univariate, single-site forecasting problem and did not analyse which models performed best from a physical perspective. As a result, it does not address the spatial, temporal, vertical, and dust-related structure present in global reanalysis products such as OpenMARS.

More recent work has started to frame Martian atmospheric prediction as a broader data-driven modelling problem. Roy~\textit{et al.}~\cite{roy2026foundation_martian_atmosphere} discuss the design landscape for a Mars Atmospheric Foundation Model, bringing together reanalysis data, atmospheric retrievals, Mars global climate models, and recent AI architectures for atmospheric physics. Their work identifies OpenMARS, spacecraft retrievals, MarsWRF, and Mars Global Climate Models as complementary sources for data-driven Martian atmospheric modelling, and discusses downstream applications including dust storms, frontal systems, low-level jets, water-ice clouds, surface-pressure forecasting, downscaling, and reanalysis--observation fusion. They also report initial experiments with Mars-adapted architectures, including a GraphCast-style graph neural model, a vision-transformer-based model, and a spherical neural operator.

This recent direction is closely related to MaGMA, but the focus is different. Foundation-model work aims to define a broad modelling agenda for the Martian atmosphere across datasets, tasks, and architectures. MaGMA focuses on a more specific question: how to transform OpenMARS fields into reusable graph-structured learning objects for multi-horizon forecasting. By representing local atmospheric patches, recent histories, vertical information, spatial neighbours, temporal links, and dynamically similar states together, MaGMA addresses a gap between single-site ML forecasting and large-scale foundation-model ambitions for Mars.

\section{OpenMARS Dataset}
\label{sec:dataset}

This study uses the publicly available Open access to Mars Assimilated Remote Soundings (OpenMARS) continuous MY28--35 reanalysis product~\cite{holmes20,streeter24}. OpenMARS combines spacecraft observations with a Mars Global Circulation Model (GCM) to provide a global reference database of Martian surface and atmospheric properties. The specific product used here is organised by Martian Year from MY28 to MY35 and begins at sol 235 of MY28. The forecasting experiments in this study use the MY28--MY34 subset described below.

For the graph training experiments in this paper, we use year-specific OpenMARS fields drawn from MY28-MY34. MY28 is used for model training and validation, while MY29, MY30, MY31, MY32, and MY34 are used for cross-year evaluation. This design separates the full temporal coverage of the OpenMARS product from the subset used in the present forecasting experiments.

Table~\ref{tab:mars_years} summarises the Martian years used in the forecasting experiments and their corresponding approximate Earth-date ranges. A complete Martian year is approximately 668.6 sols. The OpenMARS MY28 record used here begins part-way through MY28, at sol 235 ($L_s \approx 111^\circ$), rather than at the beginning of the Martian year.

\begin{table}[ht]
\centering
\caption{Martian years used in the MaGMA experiments and their approximate Earth-date coverage. MY28 begins at sol 235 in the OpenMARS product used here; the remaining evaluation years span approximately complete Martian years.}
\label{tab:mars_years}
\begin{tabular}{llll}
\toprule
Martian Year & Experimental role & Earth-date coverage & Approx. sols \\
\midrule
MY28 & Training/validation & 24 Sep 2006--8 Dec 2007 & 433 \\
MY29 & Evaluation & 9 Dec 2007--25 Oct 2009 & 669 \\
MY30 & Evaluation & 26 Oct 2009--12 Sep 2011 & 669 \\
MY31 & Evaluation & 13 Sep 2011--30 Jul 2013 & 669 \\
MY32 & Evaluation & 31 Jul 2013--17 Jun 2015 & 669 \\
MY34 & Evaluation & 5 May 2017--22 Mar 2019 & 669 \\
\bottomrule
\end{tabular}
\end{table}

OpenMARS assimilates spacecraft observations of temperature, dust, and water vapour from the Thermal Emission Spectrometer instrument~\cite{smith00,smith02,smith04} aboard NASA's Mars Global Surveyor spacecraft, as well as temperature and dust column optical depth from the Mars Climate Sounder instrument~\cite{kleinbohl17,kleinbohl20} aboard NASA's Mars Reconnaissance Orbiter spacecraft. These observations are combined with the Mars GCM using the Analysis Correction scheme~\cite{lorenc91}, which applies successive corrections to relevant variables and weights observations over short temporal and spatial windows to avoid instabilities caused by abrupt changes in the simulation state. Further details of the OpenMARS product are provided in~\cite{holmes20}.

The complete OpenMARS product is a multidimensional atmospheric dataset rather than a simple local time series. It contains gridded fields with 72 longitude points, 36 latitude points, 35 vertical levels, and 360 time outputs per file. Surface variables are stored over longitude, latitude, and time, while atmospheric variables are stored over longitude, latitude, vertical level, and time. Dust optical depth is represented as a column-integrated field over longitude, latitude, and time. Individual files cover 30 sols, and the dataset is provided in Network Common Data Form version 4 (NetCDF4) format. NetCDF4 is a self-describing, array-oriented scientific data format widely used for multidimensional geophysical and atmospheric datasets. It is well suited to reanalysis products because variables are stored together with their dimensions, coordinates, units, and metadata. This allows longitude-latitude-time and longitude-latitude-vertical-level-time fields to remain explicitly structured during processing, and not flattened into tabular records.

For this study, we use the OpenMARS reanalysis fields covering MY 28--35. The variables used to construct the learning representation are surface pressure (\texttt{ps}), surface temperature (\texttt{tsurf}), dust column optical depth (\texttt{dustcol}), zonal wind (\texttt{u}), meridional wind (\texttt{v}), and atmospheric temperature (\texttt{temp}). These variables were selected to represent the principal meteorological state fields available in the standard OpenMARS reanalysis relevant to the forecasting task, rather than through a separate feature-selection procedure. Surface pressure and surface temperature describe the near-surface thermodynamic state, dust column optical depth represents atmospheric dust loading, atmospheric temperature represents the thermal structure of the atmosphere, and the zonal and meridional wind components describe horizontal circulation. The remaining surface field in the standard OpenMARS product, surface CO$_2$ ice (\texttt{co2ice}), was not included because the present forecasting task focuses on the evolving atmospheric state rather than surface condensed CO$_2$ mass.

\begin{table}[h]
    \centering
    \caption{Variables used from the OpenMARS dataset.}
    \label{tab:var}
    {\scriptsize
    \begin{tabular}{|c|c|c|}
   
        \toprule
        Variable & Description & Unit\\
        \midrule
        tsurf & Surface temperature & Kelvin \\
        ps & Surface pressure & Pascals \\
        u & Near-surface ($\sim$4\,m) zonal  wind & m/s \\
        v & Near-surface ($\sim$4\,m) meridional wind & m/s \\
        dustcol & Dust column optical depth at visible wavelength & No unit \\
        temp & Atmospheric temperature at $\sim$20\,km altitude & Kelvin \\
        \bottomrule
    \end{tabular}}
\end{table}

\begin{table*}[htbp]
\centering
\footnotesize
\caption{Illustrative rows from the cleaned local time-series representation used in earlier work.}
\label{tab:head}
\begin{tabular}{|l|l|l|l|l|l|l|l|l|}
\hline
\textbf{Time}            & \textbf{Tsurf} & \textbf{Psurf} & \textbf{Cloud} & \textbf{Vapour} & \textbf{u\_wind} & \textbf{v\_wind} & \textbf{Dust} & \textbf{Temp} \\ \hline
1998-07-15 21:23:39 & 264.042 & 721.113 & 0.092 & 0.027 & -7.451 & 8.604 & 0.428 & 179.686 \\ \hline
1998-07-15 23:26:53 & 274.736 & 705.090 & 0.145 & 0.026 & -7.053 & 4.934 & 0.427 & 174.502 \\ \hline
1998-07-16 01:30:07 & 265.939 & 700.691 & 0.105 & 0.026 & -6.825 & -0.063 & 0.427 & 173.429 \\ \hline
1998-07-16 03:33:21 & 238.624 & 697.252 & 0.134 & 0.025 & -5.373 & -4.048 & 0.426 & 173.556 \\ \hline
1998-07-16 05:36:35 & 213.634 & 717.146 & 0.139 & 0.026 & -3.899 & -3.133 & 0.426 & 174.789 \\ \hline
\end{tabular}
\end{table*}

Table~\ref{tab:var} summarises the OpenMARS variables used in the study.
Table~\ref{tab:head} shows an illustrative sample of the earlier cleaned local time-series representation; it is included for comparison and is not the year-specific graph input used in the MaGMA experiments. The temporal resolution is two hours, giving 88,560 data points for each variable over the full period. The sampled grid point corresponds to the 5$^\circ$ gridbox centred at 2.5$^\circ$N, 135$^\circ$E, approximately matching the model gridbox over the InSight landing site~\cite{banfield2020atmosphere_of_mars_nature}. This local view is useful for benchmarking conventional time-series forecasting models, but it does not retain the full spatial and vertical structure of the OpenMARS fields.

MaGMA therefore uses OpenMARS as a gridded multidimensional reanalysis product rather than reducing it to a single-site sequence. The data engineering pipeline transforms the atmospheric fields into graph-structured learning objects for multi-horizon forecasting. In this representation, local latitude--longitude patches become graph nodes, temporal windows provide recent atmospheric histories, vertical fields provide column information, and graph edges encode spatial, temporal, and atmospheric-state relationships. The next section introduces the MaGMA framework and describes how this transformation supports graph-based Martian atmospheric forecasting.

\section{MaGMA: Martian Graph-based Multi-horizon Atmospheric Forecasting}
\label{sec:magma}

\begin{figure*}[ht]
\centering
\includegraphics[width=0.95\textwidth]{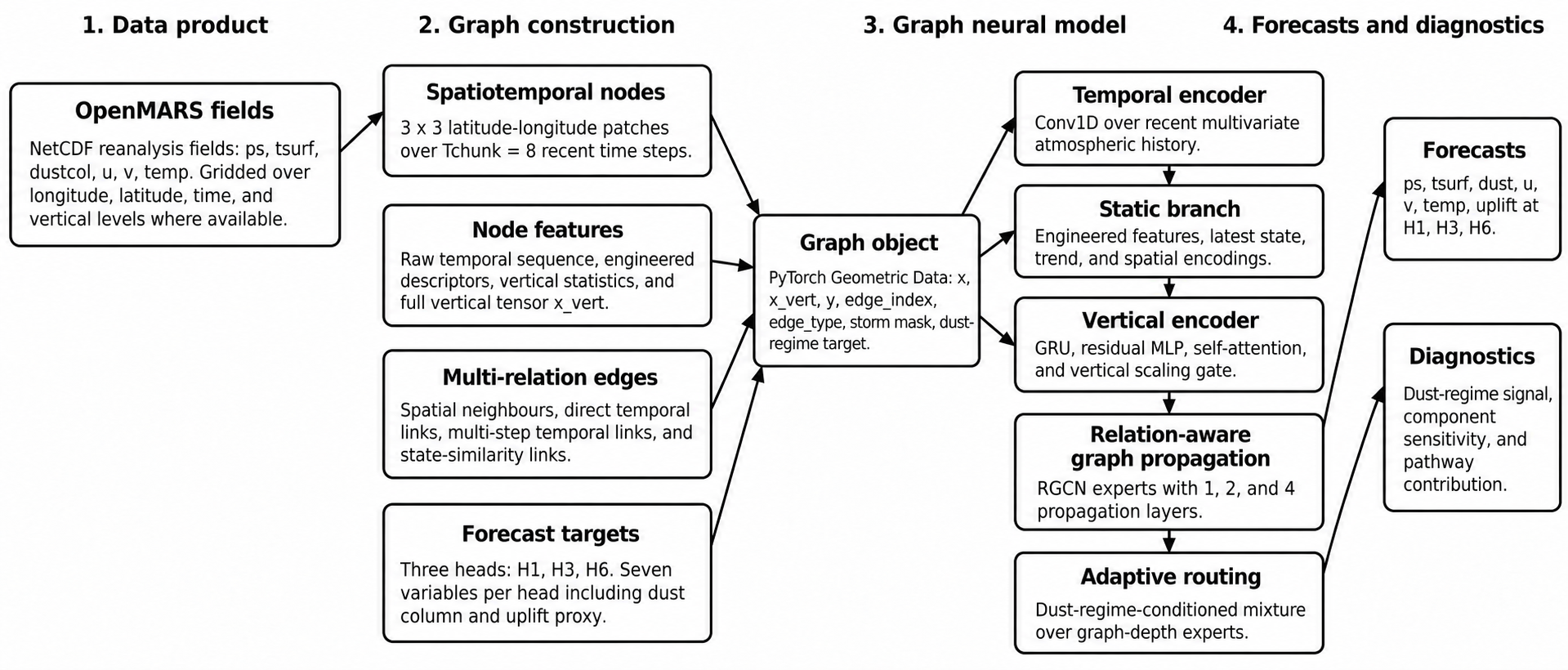}
\caption{Overview of the MaGMA workflow. OpenMARS gridded atmospheric fields are transformed into graph-structured learning objects by partitioning the data into local spatiotemporal patches, constructing node features from recent temporal history, engineered descriptors, and vertical column information, and connecting nodes through spatial, temporal, multi-step temporal, and atmospheric-state similarity edges. During graph construction, each node is also assigned multi-horizon targets corresponding to H1, H3, and H6. The resulting graph object is processed by a neural forecasting architecture that combines temporal encoding, static and engineered features, vertical column encoding, relation-aware graph propagation, adaptive routing, and horizon-specific residual forecast heads. The model predicts surface pressure, surface temperature, dust column, zonal wind, meridional wind, atmospheric temperature, and uplift at each horizon.}
\label{fig:magma_framework}
\end{figure*}

Figure~\ref{fig:magma_framework} summarises the overall MaGMA workflow, from OpenMARS field extraction and graph construction with multi-horizon target assignment through to relation-aware forecasting and diagnostic outputs.

MaGMA is a graph-based learning framework that transforms gridded Martian atmospheric reanalysis data into structured inputs for multi-horizon forecasting. The framework treats data engineering as part of the modelling problem: rather than flattening atmospheric fields into independent time series or feature vectors, it preserves spatial, temporal, vertical, and atmospheric-state relationships in a graph representation that can be processed by a graph neural network.

The MaGMA workflow has two main stages. The first stage constructs graph-structured learning objects from the OpenMARS atmospheric fields. Continuous atmospheric time series are segmented into fixed temporal windows so that each node captures the recent evolution of a local atmospheric patch. The spatial grid is partitioned into latitude--longitude regions, allowing each node to represent both a local area on the Martian surface and a short atmospheric history. The representation is enriched with engineered features, including wind magnitude, transformed dust variables, geographic encodings, seasonal indicators, and vertical atmospheric statistics. Together, these features expose physically meaningful signals while keeping the learning representation compact and reusable.

The second stage applies a neural forecasting architecture to the resulting graph. The model combines three complementary information pathways. A temporal encoder captures short-term atmospheric evolution from recent multivariate sequences. A vertical encoder processes atmospheric column information, allowing the model to incorporate structure across vertical levels. A static and engineered feature branch represents the latest atmospheric state, spatial context, and derived physical descriptors. These representations are fused before relation-aware graph propagation.

Graph propagation allows the model to exchange information across multiple types of relationships encoded in the graph, including spatial neighbourhoods, temporal continuity, longer temporal dependencies, and dynamically similar atmospheric states. Rather than using a single fixed propagation depth, MaGMA uses multiple graph propagation pathways and an adaptive routing mechanism that selects how strongly each pathway contributes for each node. This design allows the model to adjust the scale of relational information used for different atmospheric states.

The final component of MaGMA consists of multi-horizon forecast heads. These heads predict future atmospheric conditions at short, medium, and longer horizons, including surface pressure, surface temperature, dust column, zonal wind, meridional wind, atmospheric temperature, and a derived uplift indicator. The architecture therefore supports both forecasting and diagnostic analysis: it can evaluate predictive skill while also examining how temporal, engineered, vertical, and graph-based information pathways contribute to performance.

The following sections detail the two main components of this framework. Section~\ref{sec:graph} describes the graph construction process, including node features, edge relations, and forecasting targets. Section~\ref{sec:gnn} then presents the graph neural forecasting architecture used to process these graph objects.

\section{Graph Construction}
\label{sec:graph}
This section describes how the multidimensional OpenMARS reanalysis fields are transformed into supervised graph-learning objects for MaGMA. We first define the local atmospheric nodes, then describe the node features, edge relations, and multi-horizon forecasting targets that together form the graph representation used by the neural forecasting model.

\subsection{From Gridded Fields to Local Atmospheric Nodes}
The graph construction stage changes the multidimensional OpenMARS fields into supervised graph-learning objects. This is not only a preprocessing step, it defines what each node represents in the graph, what information is available to the model, how nodes are related to each other, and what future states the model is trained to predict. The challenge is that the original reanalysis fields contain spatial grids, temporal evolution, vertical structure, and variables with different physical meanings. MaGMA addresses this by partitioning the fields into local spatiotemporal chunks, representing each chunk as a graph node, adding relation-specific edges between nodes, and attaching multi-horizon forecasting targets to each node. The forecasting targets are described in this section because they are constructed at the same time as the graph nodes: each node is paired with future atmospheric states at different horizons, allowing the final graph object to contain both the input representation and the supervised prediction targets.

Spatially, the global grid is divided into latitude--longitude patches of size $3 \times 3$. The native OpenMARS horizontal resolution is $5^\circ \times 5^\circ$, so each graph node represents a $15^\circ \times 15^\circ$ atmospheric region. This patch size was chosen as a practical compromise between retaining regional spatial structure and controlling the size of the spatiotemporal graph. Operating at the native grid resolution would produce 2,592 spatial nodes per graph time step, whereas the $3 \times 3$ aggregation reduces this to 288 nodes while preserving global coverage and local neighbourhood relationships. Since the OpenMARS grid contains 36 latitude points and 72 longitude points, the patches tile the grid exactly $(\frac{36}{3} \times \frac{72}{3} = 12 \times 24 = 288)$. The $3 \times 3$ configuration is therefore treated as a pragmatic representation scale rather than an empirically optimised spatial resolution; systematic evaluation of alternative patch sizes is left for future work.

For each temporal window and latitude--longitude patch, MaGMA creates one graph node. Variables with vertical structure retain all 35 model levels, allowing the node representation to preserve column information as well as surface and near-surface conditions. Each node therefore encodes an atmospheric state consisting of an eight-step temporal history over a $3 \times 3$ spatial footprint, enriched with engineered atmospheric descriptors and associated with multi-horizon forecasting targets.

This transformation turns the original gridded reanalysis fields into a graph-structured learning representation. Nodes correspond to local atmospheric states, while the edges introduced in later subsections encode spatial, temporal, and atmospheric-state similarity relationships.

\subsection{Node Feature Engineering}
For each node, we construct features that capture both the recent behaviour of the local atmospheric patch and its current physical state. These include raw temporal sequences, derived atmospheric quantities, vertical summaries, and the full column information used by the model.

\paragraph{Raw Temporal Sequence Features}

Each node contains temporal sequence features for the six core atmospheric variables,
$
\texttt{ps},\; \texttt{tsurf},\; \texttt{dustcol},\; \texttt{u},\; \texttt{v},\; \texttt{temp}.
$
For each variable, values are averaged over the local \(3 \times 3\) atmospheric patch and stored over the eight most recent time steps in the node window. The raw temporal component therefore contains
$6 \times 8 = 48$
sequence features per node. This sequence provides the temporal encoder with recent local atmospheric evolution before each predicted horizon.

\paragraph{Engineered Atmospheric Features}

Additional engineered features were added from the final step of each node window. Using only the final observed step preserves causal structure because no future information is included in the input representation.

The engineered features include current wind speed, logarithmic dust loading, square-root transformed dust loading, temperature anomaly, pressure anomaly, normalized latitude patch index, normalized longitude patch index, a binary dust-regime flag, and a current uplift proxy. Wind speed was computed from the zonal and meridional wind components:
$
U = \sqrt{u^2 + v^2}.
$

Dust loading was represented using both logarithmic and square-root transformations to help capture nonlinear variation in the dust column field. The binary dust-regime flag was set to one when the local dust column exceeded a fixed threshold and zero otherwise. This flag was also used as a weak auxiliary target for the dust-regime classification head.

A simple uplift proxy was used from wind, temperature, and dust loading. This proxy is not a directly validated core variable of dust lifting. Instead, it is used as a diagnostic quantity to capture dust-active atmospheric conditions.

\paragraph{Vertical Summary Statistics}

In addition to the raw temporal and engineered features, vertical summary statistics were calculated for variables with vertical structure. These statistics were extracted from the final time step of the node window and included the lower-column mean, upper-column mean, vertical standard deviation, and lower-to-upper column contrast. These features provide compact information about the vertical structure of the atmosphere while keeping the flat node feature vector manageable.

For variables without an explicit vertical dimension, zero-valued summary statistics were inserted so that the feature schema remained consistent across all variables.

\paragraph{Full Vertical Column Tensor}

During graph construction, the full vertical column profile was also kept and stored in a tensor denoted \(X_{\mathrm{vert}}\). This was designed for the vertical encoder in the neural architecture. The vertical input variables are:
$
\texttt{temp},\; \texttt{u},\; \texttt{v},\; \texttt{dustcol},\; \texttt{ps}.
$
Surface temperature \texttt{tsurf} is included in the raw temporal sequence and forecasting targets, but not in \(X_{\mathrm{vert}}\), because it is a surface field rather than a vertical atmospheric profile. Vertical thermal structure is represented by \texttt{temp}. Variables without a native vertical dimension, such as \texttt{ps} and \texttt{dustcol}, are broadcast across vertical levels so that all vertical input channels share the same tensor shape.

The vertical profile was extracted at the final step of the node window and averaged over the local latitude--longitude atmospheric patch. For variables that did not have an explicit vertical dimension, the scalar patch mean was broadcast across all vertical levels. This ensured that all variables had a consistent tensor shape:
$
X_{\mathrm{vert}} \in \mathbb{R}^{N \times Z \times V},
$
where \(N\) is the number of graph nodes, \(Z\) is the number of vertical levels, and \(V\) is the number of vertical input variables.

The flat node feature matrix and vertical tensor were normalized separately using global means and standard deviations computed across the graph.

\subsection{Graph Edge Construction}

Once nodes are created, edges are added to define the relational structure over which graph message passing operates. The purpose of these edges is to make explicit the main relationships that may carry forecasting information: local spatial adjacency, temporal evolution, short-range temporal continuity, and similarity between atmospheric states. Four edge relation types were therefore used: local spatial edges, direct temporal edges, multi-step temporal edges, and state-similarity edges.

\begin{itemize}
    \item \textit{Local spatial edges} connect neighbouring latitude--longitude patches within the same time window. An eight-neighbourhood was used around each node, excluding the node itself. Longitude wrapping was included so that the global longitudinal boundary remained connected.

    \item \textit{Direct temporal edges} connect a node at graph time \(t\) to the same spatial patch at graph time \(t+1\). These edges encode forward temporal evolution of the local atmospheric state.

    \item \textit{Multi-step temporal edges} connect a node at graph time \(t\) to the same spatial patch up to two graph-time steps ahead (\(g \leq 2\)). The maximum gap was limited to two in order to provide short temporal skip connections while keeping the graph temporally local and avoiding the rapid growth in connectivity that would result from adding longer-range links. Since each graph-time node represents an eight-step OpenMARS window, these edges provide additional short-range temporal context without attempting to encode long-range atmospheric evolution directly through the graph topology. The value \(g \leq 2\) is therefore a pragmatic locality constraint rather than an empirically optimised temporal scale.

    \item \textit{State-similarity edges} connect atmospheric patches within the same time slice using \(k\)-nearest neighbours over normalized node features. Cosine distance was used as the similarity metric, with
    $
    K_{\mathrm{sim}} = 5.
    $
    A small neighbourhood of five was used to keep the state-similarity relation sparse and to allow a limited number of dynamically similar but geographically distant regions to exchange information without allowing feature-space connectivity to dominate the physically defined spatial and temporal graph structure. The value \(K_{\mathrm{sim}}=5\) is therefore treated as a pragmatic sparsity choice rather than an empirically optimised neighbour count. This interpretation is also consistent with the component sensitivity analysis, in which removing similarity edges produces only a small change in aggregate forecasting performance.
\end{itemize}

The final graph connectivity was stored using \texttt{edge\_index}, while relation labels were stored using \texttt{edge\_type}. The graph was then packaged as a PyTorch Geometric data object containing the node features, vertical tensor, target matrix, graph connectivity, edge relation labels, storm mask, and auxiliary dust-regime target.

\subsection{Forecasting Targets}

The graph construction process assigns three supervised forecasting targets to each node, corresponding to the forecast horizons H1, H3, and H6. These are not ensemble members. They are three separate future offsets, each represented by a separate seven-variable target vector.
Formally, the horizon index is:
$
H \in \{1, 3, 6\}.
$

Each horizon is defined as a future graph-chunk offset of \(H \times T_{\mathrm{chunk}}\) relative to the current node window. Since \(T_{\mathrm{chunk}} = 8\), H1, H3, and H6 correspond to target windows shifted by 8, 24, and 48 raw OpenMARS time indices.

The first six targets correspond to future values of the core atmospheric variables. The seventh target is a derived uplift proxy computed from future wind speed, positive temperature change, and future dust loading:
$
\texttt{uplift}
=
U_{\mathrm{future}}
\times
\max(\texttt{temp}_{\mathrm{future}} - \texttt{temp}_{\mathrm{current}}, 0)
\times
\log(\texttt{dust}_{\mathrm{future}} + \epsilon).
$

For each node, the three seven-variable target vectors were concatenated into a 21-dimensional target vector:
$
y =
[y_{H1}, y_{H3}, y_{H6}]
\in \mathbb{R}^{21}.
$

Rows where the required future target window exceeded the available time range were assigned missing values and excluded from evaluation through finite-value masking. The final target matrix was normalized using global target means and standard deviations before training.

\section{Graph Neural Architecture for Multi-Horizon Forecasting}
\label{sec:gnn}
This section describes the neural forecasting architecture used to process the graph objects constructed in Section~\ref{sec:graph}. Each node provides the model with recent local atmospheric history, engineered descriptors of the current state, vertical column information, and graph relations to other atmospheric patches. The architecture first encodes these node-level inputs into a shared representation, then applies relation-aware graph propagation so that each forecast can use spatial neighbours, temporal links, and dynamically similar atmospheric states. The routed graph representation is finally passed to horizon-specific residual forecast heads, which predict departures from the latest observed state for H1, H3, and H6.

\paragraph{Model Overview}

The proposed model is a graph neural network designed to operate on the spatiotemporal graph objects described in Section~\ref{sec:graph}. Each graph node represents a local Martian atmospheric patch and contains recent temporal evolution, engineered atmospheric descriptors, vertical column information, and graph relations to other atmospheric patches.

The model combines three input pathways. First, a temporal encoder processes the recent multivariate atmospheric history of each node. Second, a static feature branch processes engineered atmospheric features, spatial encodings, the final observed state, and the most recent local trend. Third, a vertical column encoder processes the full atmospheric column tensor for selected variables. These representations are fused into a shared node embedding before relation-aware graph propagation is applied.

Graph propagation is performed using relation-aware graph convolution over the edge types generated during graph construction. These include spatial, temporal, multi-step temporal, and similarity-based relations. Rather than using a single fixed propagation depth, the model uses multiple graph propagation pathways of different depths and combines them through a learned routing mechanism. The routing module is conditioned on the fused node representation and an auxiliary dust-regime signal.

The model produces three multi-horizon forecasts corresponding to H1, H3, and H6. Each horizon predicts seven target variables: surface pressure, surface temperature, dust column, zonal wind, meridional wind, atmospheric temperature, and uplift. Forecasting is formulated as a residual correction around a persistence baseline constructed from the final observed atmospheric state. The overall workflow is summarised in Figure~\ref{fig:magma_framework}, while the following subsections describe the architecture components in detail.

\paragraph{Temporal Encoder}

The temporal branch receives the raw sequence component of each node. For every local atmospheric patch, six variables are provided over eight recent time steps:
$
\texttt{ps},\; \texttt{tsurf},\; \texttt{dustcol},\; \texttt{u},\; \texttt{v},\; \texttt{temp}.
$

This gives a temporal input of size \(6 \times 8\) for each node. The purpose of this branch is to learn short-term local atmospheric evolution before the prediction horizon.

The temporal encoder uses one-dimensional convolutions over the sequence. In implementation, two convolutional layers with Gaussian Error Linear Unit (GELU) activations are followed by adaptive average pooling across the temporal dimension. The resulting temporal embedding is then projected into the shared hidden dimension used by the rest of the model. This pathway is intended to capture short-term patterns in dust loading, pressure, temperature, and wind evolution within each local atmospheric patch.

\paragraph{Static and Engineered Feature Branch}

The second branch processes the non-sequential node features. These include engineered atmospheric descriptors, spatial patch encodings, transformed dust variables, wind-speed information, anomaly features, uplift-proxy information, and compact vertical summary statistics.

In addition to these precomputed features, the model explicitly extracts two short-term descriptors from the raw temporal sequence: the final observed atmospheric state and the most recent one-step trend. The final state gives the network direct access to the latest observed values, while the trend captures the most recent local direction of change. These descriptors are concatenated with the engineered feature vector and passed through a feed-forward projection with GELU activation and layer normalization.

This branch provides the model with immediately available local state information and supports the residual forecasting design by exposing the latest atmospheric state from which future deviations are learned.

\paragraph{Vertical Column Encoder}

The third branch processes the full vertical column tensor, denoted \(x_{\mathrm{vert}}\). This tensor contains vertical profiles for selected variables at the final time step of the node window. In this implementation, the vertical input variables are:
$
\texttt{temp},\; \texttt{u},\; \texttt{v},\; \texttt{dustcol},\; \texttt{ps}.
$

Variables with explicit vertical structure are represented across the vertical levels available in the OpenMARS data. Variables without a native vertical dimension are broadcast across levels during graph construction to maintain a consistent tensor shape.

The vertical encoder first projects the input variables at each level into a hidden representation. A gated recurrent unit is then applied along the vertical dimension, allowing the model to process the atmospheric column as an ordered vertical sequence. The resulting representation is passed through a residual multilayer perceptron and a self-attention layer. The attention layer allows the model to reweight interactions across vertical levels before the column representation is pooled and projected into the shared hidden dimension.

A learned scalar gate modulates the contribution of the vertical pathway during fusion. The fused pre-graph representation can be summarized as:
$
h_{\mathrm{base}}
=
h_{\mathrm{temp}}
+
h_{\mathrm{static}}
+
\gamma h_{\mathrm{vert}},
$

where \(h_{\mathrm{temp}}\), \(h_{\mathrm{static}}\), and \(h_{\mathrm{vert}}\) are the temporal, static, and vertical embeddings, respectively, and \(\gamma\) is the learned vertical scaling factor.

This pathway gives the model access to vertical atmospheric structure while still allowing later analysis to determine how strongly the trained model depends on that information.

\paragraph{Auxiliary Dust-Regime Head}

The fused node representation is passed to an auxiliary dust-regime classification head. This head predicts a weak binary dust-active label derived from the dust-threshold flag used during graph construction. In the manuscript, this output is referred to as the auxiliary dust-regime signal rather than as an independently defined physical variable.

The auxiliary head has two roles. First, it provides an additional training signal associated with dust-active atmospheric states. Second, its predicted probability is supplied to the routing module, allowing graph propagation to be conditioned on the inferred local dust regime.

The auxiliary dust-regime signal is treated as an auxiliary regime indicator rather than as an independently validated physical dust storm classifier. Its behaviour is evaluated separately in Section~\ref{sec:results}, where it is used to compare regular-year behaviour with MY34 global dust storm conditions.

\paragraph{Relational Graph Propagation}

After feature fusion, the model applies relation-aware graph propagation over the multi-relation graph. The graph contains four relation types: local spatial edges, direct temporal edges, multi-step temporal edges, and state-similarity edges. These relation labels are used by relational graph convolution layers so that messages can be conditioned on the type of connection between nodes.

Each graph propagation block applies relational graph convolution, incorporates edge-type information through learned relation embeddings, and then applies nonlinear activation, layer normalization, dropout, and a residual connection. The residual connection helps stabilize graph propagation and reduces the risk of excessive smoothing when multiple graph convolution layers are used.

This graph propagation step allows information exchange between physically adjacent atmospheric patches, temporally connected patches, short multi-step temporal neighbours, and dynamically similar regions within the same time slice. The model therefore does not forecast each atmospheric patch independently; instead, it learns from the relational structure encoded during graph construction.

\paragraph{Adaptive Routing over Propagation Depth}

The model uses three graph propagation pathways with different depths, implemented using relational graph convolution layers (RGCNConv) from PyTorch Geometric: 
$
1 \times \mathrm{RGCNConv}, \qquad
2 \times \mathrm{RGCNConv}, \qquad
4 \times \mathrm{RGCNConv}.
$

These pathways act as graph experts with different effective receptive fields. The shallow pathway preserves more local information, while the deeper pathways allow information to travel through wider relational neighbourhoods.

A learned router combines the three expert outputs. The router receives the fused node representation together with the predicted auxiliary dust-regime probability and produces a softmax weighting over the graph experts. The final graph representation is a weighted mixture:
$
h_{\mathrm{mix}}
=
g_1 h_1
+
g_2 h_2
+
g_3 h_3,
$

where \(h_1\), \(h_2\), and \(h_3\) are the outputs of the shallow, medium, and deeper graph propagation pathways, and \(g_1\), \(g_2\), and \(g_3\) are the learned routing weights.

During training, a temperature parameter is used in the router softmax and is gradually annealed. This encourages smoother expert weighting early in training and more confident routing later. The routing mechanism allows the model to adapt the effective graph propagation depth according to the local atmospheric representation and auxiliary dust-regime signal.

\paragraph{Multi-Horizon Residual Forecast Heads}

The final mixed graph representation is passed to three separate forecast heads, one for each forecast horizon:
$H1,\; H3,\; H6.$

Each head predicts seven residual values corresponding to:
$
\texttt{ps},\; \texttt{tsurf},\; \texttt{dust},\; \texttt{u},\; \texttt{v},\; \texttt{temp},\; \texttt{uplift}.
$

The model uses residual forecasting around a persistence baseline rather than predicting the full future state directly. The persistence baseline is constructed from the final observed state of the six core atmospheric variables. Since uplift is a derived target rather than a directly observed sequence variable, its persistence baseline is set to zero. The model therefore learns deviations from the most recent atmospheric state:
$
\hat{y}_H = b + r_H,
$
where \(b\) is the persistence baseline and \(r_H\) is the learned residual for horizon \(H\).

This formulation is useful because many atmospheric variables exhibit short-term persistence. The network can retain this baseline behaviour while learning departures associated with dust evolution, wind-field changes, thermal variation, and longer-horizon atmospheric dynamics.

The three forecast heads produce a combined 21-dimensional output per node, matching the multi-horizon target vector.

\paragraph{Numerical Stabilization and Regularization}

Several stabilization steps were used during model implementation. Tensor values were sanitized using finite-value replacement and clipping to prevent invalid numerical values from propagating through the network. Layer normalization and dropout were used in the static, vertical, and graph propagation branches. Gradient clipping was applied during optimization.

A small amount of Gaussian noise was injected into the final temporal state features during training as a lightweight regularization strategy. This was used to reduce over-reliance on exact last-step values and to improve robustness around the persistence baseline.

Overall, the architecture combines temporal sequence modelling, engineered atmospheric descriptors, vertical column encoding, relation-aware graph propagation, auxiliary dust-regime conditioning, adaptive routing, and residual multi-horizon forecasting. This design supports both predictive evaluation and diagnostic analysis of the information pathways used for Martian dust and atmospheric forecasting.

\section{Experiments}
\label{sec:experimetns}
This section sets out the experimental setup and then presents the results. We describe how the model was trained on MY28, evaluated on later Martian years including MY34 global dust storm year, and compared with both classical and temporal forecasting baselines.

\subsection{Training and Evaluation Strategy}
In this subsection, we describe how we train MaGMA, evaluate it across Martian years, compare it with baseline models, and measure forecasting performance.

\paragraph{Scalable Subgraph Training}

The graph construction process produces large atmospheric graphs, making full-graph graphics processing unit (GPU) training impractical within the available accelerator memory. Graph construction, preprocessing, and storage of the complete graph were therefore performed in CPU memory, while model training used time-batched induced subgraphs transferred to the GPU. The experiments were conducted in a Google Colab environment using an NVIDIA A100 GPU. The host CPU configuration was provided dynamically by the Colab runtime and was not treated as a controlled experimental variable. This CPU--GPU division allowed relation-aware graph message passing to be used without transferring the complete atmospheric graph to GPU memory at once.

Before training, the graph tensors were converted to contiguous CPU tensors. This included the node feature matrix, vertical tensor, target matrix, edge index, edge type labels, storm mask, and auxiliary dust-regime target. Edge indices and edge relation labels were validated before training to ensure that node identifiers were within valid bounds and that relation labels were non-negative.

Training batches were created by grouping nodes according to their graph-time index. For each batch, a set of consecutive time slices was selected and the corresponding induced subgraph was extracted from the original graph. The subgraph was then relabelled locally and moved to GPU for training. The default graph-time batch size was:
$
batch_t = 8.
$

Each graph-time slice contains 288 spatial nodes. Consequently, a full eight-slice training batch contains up to \(8 \times 288 = 2{,}304\) node entries before induced-subgraph extraction. The number of edges in each induced subgraph varies according to temporal position and the spatial, temporal, multi-step, and state-similarity relations retained within the selected time window.

Within each time batch, dust-active nodes were oversampled to increase the representation of dust-active atmospheric states during training. The target storm ratio was set to:
$
r_{\mathrm{storm}} = 0.25.
$
This was used to reduce the dominance of calm atmospheric states and to give the model more exposure to dust-active examples.

Mixed precision training was used when Compute Unified Device Architecture (CUDA) was available. Gradient accumulation was also used to control memory requirements while increasing the effective optimization batch size. In this study, the accumulation step was:
$
grad_{\mathrm{accum}} = 2.
$

Gradient clipping was applied during optimization to improve numerical stability.

\paragraph{Train--Validation Split and Cross-Year Evaluation}
The GNN was trained using a time-based train--validation split on MY28. Nodes were split according to graph-time index rather than random node sampling. This reduces leakage between adjacent temporal windows and provides a more appropriate validation setting for forecasting. The first 80\% of unique graph-time indices in MY28 were used for training, and the remaining 20\% were used for validation.

The final GNN was trained on the MY28 graph and evaluated on year-specific graphs for MY29, MY30, MY31, MY32, and MY34. This setup evaluates whether the model trained on one Martian year can generalize to subsequent regular years and to the MY34 global dust storm regime.

During evaluation, inference was also performed using time-batched induced subgraphs. Rows with missing future targets, which occur when the required forecast horizon extends beyond the available time range, were excluded from metric calculations using finite-value masking.

\paragraph{Training Objective}

The training objective combines multi-horizon regression with auxiliary dust-regime learning and several regularization terms. For each graph node, the model predicts three seven-dimensional output vectors corresponding to H1, H3, and H6. These are compared with the corresponding normalized target vectors.

The main regression loss is a weighted mean squared error applied separately to each forecast horizon. The target weights were:
$
w = [1,\;1,\;4,\;1,\;1,\;1,\;5].
$

These weights place greater emphasis on dust column and uplift relative to the other variables. This reflects the focus of the study on dust-column forecasting and dust-active atmospheric behaviour.

Dust-active nodes were also given higher regression weight through the storm mask. This encouraged the model to learn from dust-active states, which are important for global dust storm analysis but do not dominate the dataset.

The auxiliary dust-regime head was trained using binary cross-entropy with logits against the weak dust-regime target created during graph construction. The auxiliary dust-regime loss coefficient was:
$
\lambda_{\mathrm{dust}} = 0.15.
$

A coupled correlation loss was also included. This term compares the correlation structure of predicted outputs with the correlation structure of the target outputs for valid rows in the batch. It was included to encourage the model to preserve relationships between predicted atmospheric variables, not only individual pointwise values. The coefficient for this term was:
$
\lambda_{\mathrm{corr}} = 0.03.
$

A small entropy regularization term was applied to the router gates to reduce premature collapse of the adaptive routing mechanism. The coefficient was:
$
\lambda_{\mathrm{gate}} = 10^{-3}.
$

A small output smoothness penalty was also included as a numerical regularizer. This term discourages abrupt prediction changes within the processed batch ordering, but it is not treated here as a physical smoothness constraint. The coefficient for this term was:
$
\lambda_{\mathrm{smooth}} = 0.01.
$

The total objective can be summarized as:
$
\mathcal{L}
=
\mathcal{L}_{H1}
+
\mathcal{L}_{H3}
+
\mathcal{L}_{H6}
+
\lambda_{\mathrm{corr}}\mathcal{L}_{\mathrm{corr}}
+
\lambda_{\mathrm{dust}}\mathcal{L}_{\mathrm{dust}}
-
\lambda_{\mathrm{gate}}\mathcal{H}_{\mathrm{gate}}
+
\lambda_{\mathrm{smooth}}\mathcal{L}_{\mathrm{smooth}}.
$

\paragraph{Optimization Configuration}

The model was optimized using AdamW with weight decay. The learning rate was set to:
$
\eta = 2 \times 10^{-3}.
$

The hidden dimension was set to 256 and dropout was set to 0.05. The vertical encoder was processed in internal batches to control memory usage, with a vertical batch size of 2048 nodes. Training was run for a maximum of 150 epochs, with early stopping based on validation loss. The early-stopping patience was set to 25 epochs.

The router temperature was annealed during training. It started at 1.5 and decreased by 0.01 per epoch until reaching a lower bound of 0.7. This was used to encourage smoother expert weighting early in training and more decisive routing later.

The final model used for evaluation was the checkpoint with the best validation loss.

\paragraph{Baseline Models}

To evaluate whether the graph-based model improves dust-column forecasting, two groups of baselines were used: classical feature-based baselines and deep temporal baselines.

The classical baselines were Linear Regression and Random Forest. These models were trained using MY28 graph-node feature vectors and evaluated on MY29, MY30, MY31, MY32, and MY34. The purpose of these baselines was to compare the proposed GNN against both a simple linear model and a nonlinear feature-based ensemble model. Dust-column forecasting was used as the primary baseline comparison task.

The deep temporal baselines were LSTM, gated recurrent unit (GRU), and TCN models. These models were also trained on MY28 and evaluated on MY29, MY30, MY31, MY32, and MY34, matching the cross-year evaluation setup used for the GNN. Unlike the GNN, the temporal baselines used the raw temporal sequence component of each node and did not use the graph edge structure.

All deep temporal baselines were trained to predict the same 21-dimensional multi-horizon target vector as the GNN. The LSTM and GRU baselines used recurrent sequence encoders, while the TCN baseline used temporal convolutions. These baselines were included to test whether the proposed graph-based model provides benefit beyond temporal-only forecasting approaches.

This baseline set provides comparison against a linear classical model, a nonlinear classical ensemble, recurrent temporal models, a convolutional temporal model, and the proposed graph neural network.

\paragraph{Evaluation Metrics}

Forecasting performance was evaluated using mean absolute error, root mean squared error, and coefficient of determination:

We use standard metrics to evaluate forecasting performance: Mean Absolute Error (MAE), Root Mean Squared Error (RMSE), and coefficient of determination ($R^2$).

Metrics were computed for each Martian year, forecast horizon, and target variable. Overall metrics were also computed after masking rows containing missing or non-finite values.

Dust-column prediction was treated as the primary forecasting task because dust loading is central to global dust storm behaviour and to the scientific motivation of this study. Therefore, baseline comparison tables focus on dust-column \(R^2\) across H1, H3, and H6.

Auxiliary dust-regime classification was evaluated using accuracy, precision, recall, F1 score, and receiver operating characteristic area under the curve (ROC-AUC). These metrics were computed by comparing the predicted auxiliary dust-regime probability against the weak binary dust-regime target. The auxiliary classification analysis was used diagnostically to assess whether the regime signal behaved differently in regular years compared with MY34 global dust storm conditions.

Regional forecast skill was evaluated using regional \(R^2\). Nodes were grouped by spatial patch, and dust-column forecast performance was computed separately for each region. This allowed spatial heterogeneity in prediction skill to be assessed, especially during MY34.

Finally, inference-time component sensitivity analysis was used to estimate the importance of different model inputs and graph components. This analysis evaluated the trained model under modified inference settings in which selected components were removed or masked, including temporal sequence features, engineered/static features, vertical column inputs, full graph edges, similarity edges, temporal graph edges, and spatial graph edges. Performance degradation relative to the full model was measured using changes in overall \(R^2\) and dust H3 \(R^2\).

This component sensitivity analysis was not treated as a full retrained ablation study. Instead, it was used as a diagnostic test of the trained model's dependence on each information pathway.


\subsection{Results}
\label{sec:results}

\paragraph{Overall Multi-Year Forecasting Performance}

Table~\ref{tab:overall_generalisation} reports the overall forecasting performance across MY29, MY30, MY31, MY32, and MY34. The model shows stable performance across MY29--MY32, with overall \(R^2\) values ranging from 0.726 to 0.851. The strongest overall performance occurs in MY30, where the model achieves an MAE of 0.156, an RMSE of 0.303, and an \(R^2\) of 0.851.

MY34 shows a clear degradation in overall performance, with MAE increasing to 0.458, RMSE increasing to 0.753, and overall \(R^2\) falling to 0.363. This indicates that the model does not transfer uniformly from the MY28 training regime to the MY34 global dust storm regime. Although MY28 also contains global dust storm behaviour, the MY34 results show that exposure to one dust-storm year is not sufficient to guarantee robust generalisation to another extreme dust regime.

\begin{table}[ht]
\centering
\caption{Overall generalisation performance}
\label{tab:overall_generalisation}
\begin{tabular}{cccc}
\hline
Year & MAE & RMSE & $R^2$ \\
\hline
MY29 & 0.271 & 0.522 & 0.728 \\
MY30 & 0.156 & 0.303 & 0.851 \\
MY31 & 0.271 & 0.511 & 0.738 \\
MY32 & 0.275 & 0.524 & 0.726 \\
MY34 & 0.458 & 0.753 & 0.363 \\
\hline
\end{tabular}
\end{table}

\paragraph{Variable- and Horizon-Specific Forecast Skill}

Tables~\ref{tab:mae_horizon_variable}, \ref{tab:rmse_horizon_variable}, and \ref{tab:r2_horizon_variable} report MAE, RMSE, and \(R^2\) across forecast horizons and variables. The model performs strongly for surface pressure across all years and horizons. Dust-column forecasting also remains strong across the regular years, particularly at shorter horizons.

The \(R^2\) results show that pressure, dust, temperature, and zonal wind are generally predicted with stronger skill than meridional wind and uplift. The meridional wind component \(v\) is especially difficult under MY34, where \(R^2\) becomes negative at all horizons. Uplift is also challenging, with low or negative \(R^2\) in several cases. This is expected because uplift is a derived diagnostic proxy rather than a directly observed core atmospheric variable.
The weaker performance for meridional wind may reflect both physical and statistical factors. Zonal flow is a dominant component of the large-scale Martian circulation, while meridional wind can be more variable, regionally structured, and sensitive to transient atmospheric disturbances. This makes (v) harder to predict from the available local history and graph relations, especially under the disturbed MY34 global dust storm regime.

The MY34 results should also be interpreted carefully. The model is not being evaluated as a pre-onset global dust storm alarm. Each prediction is made from the recent atmospheric history contained in the current graph node, and the targets correspond to future offsets H1, H3, and H6. Therefore, strong MY34 dust-column performance indicates that the model can forecast the continuation and near-term evolution of elevated dust conditions once those conditions are represented in the input history. It does not by itself demonstrate that the model predicts the initial onset of the global dust storm before dust loading begins to rise.

\begin{table*}[ht]
\centering
\caption{MAE across horizons and variables}
\label{tab:mae_horizon_variable}
\begin{tabular}{ccccccccc}
\hline
Year & Horizon & ps & \texttt{tsurf} & \texttt{dust} & \texttt{u} & \texttt{v} & \texttt{temp} & \texttt{uplift} \\
\hline
MY29 & H1 & 0.052 & 0.333 & 0.079 & 0.174 & 0.482 & 0.238 & 0.497 \\
MY29 & H3 & 0.063 & 0.173 & 0.176 & 0.187 & 0.573 & 0.200 & 0.471 \\
MY29 & H6 & 0.070 & 0.184 & 0.244 & 0.212 & 0.579 & 0.215 & 0.489 \\
MY30 & H1 & 0.032 & 0.107 & 0.042 & 0.122 & 0.339 & 0.066 & 0.178 \\
MY30 & H3 & 0.047 & 0.095 & 0.086 & 0.151 & 0.434 & 0.082 & 0.232 \\
MY30 & H6 & 0.056 & 0.110 & 0.121 & 0.174 & 0.439 & 0.109 & 0.253 \\
MY31 & H1 & 0.053 & 0.333 & 0.103 & 0.164 & 0.475 & 0.233 & 0.515 \\
MY31 & H3 & 0.065 & 0.172 & 0.191 & 0.174 & 0.569 & 0.192 & 0.473 \\
MY31 & H6 & 0.073 & 0.181 & 0.246 & 0.197 & 0.579 & 0.206 & 0.497 \\
MY32 & H1 & 0.053 & 0.328 & 0.111 & 0.167 & 0.484 & 0.230 & 0.510 \\
MY32 & H3 & 0.065 & 0.169 & 0.202 & 0.181 & 0.580 & 0.190 & 0.479 \\
MY32 & H6 & 0.074 & 0.179 & 0.258 & 0.206 & 0.589 & 0.204 & 0.508 \\
MY34 & H1 & 0.058 & 0.670 & 0.108 & 0.641 & 0.741 & 0.268 & 0.392 \\
MY34 & H3 & 0.105 & 0.563 & 0.204 & 0.546 & 1.341 & 0.259 & 0.553 \\
MY34 & H6 & 0.095 & 0.565 & 0.299 & 0.537 & 1.002 & 0.307 & 0.372 \\
\hline
\end{tabular}
\end{table*}

\begin{table*}[ht]
\centering
\caption{RMSE across horizons and variables}
\label{tab:rmse_horizon_variable}
\begin{tabular}{ccccccccc}
\hline
Year & Horizon & \texttt{ps} & \texttt{tsurf} & \texttt{dust} & \texttt{u} & \texttt{v} & \texttt{temp} & \texttt{uplift} \\
\hline
MY29 & H1 & 0.070 & 0.436 & 0.163 & 0.237 & 0.679 & 0.316 & 0.929 \\
MY29 & H3 & 0.090 & 0.245 & 0.323 & 0.274 & 0.804 & 0.293 & 0.942 \\
MY29 & H6 & 0.101 & 0.282 & 0.448 & 0.322 & 0.824 & 0.338 & 0.982 \\
MY30 & H1 & 0.048 & 0.148 & 0.074 & 0.170 & 0.455 & 0.110 & 0.371 \\
MY30 & H3 & 0.077 & 0.152 & 0.129 & 0.222 & 0.597 & 0.158 & 0.457 \\
MY30 & H6 & 0.091 & 0.189 & 0.169 & 0.264 & 0.608 & 0.211 & 0.515 \\
MY31 & H1 & 0.071 & 0.432 & 0.201 & 0.217 & 0.660 & 0.305 & 0.934 \\
MY31 & H3 & 0.095 & 0.238 & 0.340 & 0.241 & 0.788 & 0.273 & 0.908 \\
MY31 & H6 & 0.108 & 0.268 & 0.437 & 0.281 & 0.816 & 0.310 & 0.961 \\
MY32 & H1 & 0.071 & 0.427 & 0.221 & 0.228 & 0.675 & 0.305 & 0.910 \\
MY32 & H3 & 0.094 & 0.240 & 0.369 & 0.264 & 0.805 & 0.282 & 0.931 \\
MY32 & H6 & 0.108 & 0.275 & 0.465 & 0.314 & 0.833 & 0.326 & 1.011 \\
MY34 & H1 & 0.081 & 0.829 & 0.351 & 0.727 & 1.056 & 0.356 & 0.643 \\
MY34 & H3 & 0.130 & 0.743 & 0.502 & 0.639 & 1.652 & 0.343 & 0.875 \\
MY34 & H6 & 0.127 & 0.748 & 0.644 & 0.639 & 1.373 & 0.414 & 0.779 \\
\hline
\end{tabular}
\end{table*}

\begin{table*}[ht]
\centering
\caption{\(R^2\) across horizons and variables}
\label{tab:r2_horizon_variable}
\begin{tabular}{ccccccccc}
\hline
Year & Horizon & \texttt{ps} & \texttt{tsurf} & \texttt{dust} & \texttt{u} & \texttt{v} & \texttt{temp} & \texttt{uplift} \\
\hline
MY29 & H1 & 0.995 & 0.810 & 0.973 & 0.944 & 0.540 & 0.900 & 0.048 \\
MY29 & H3 & 0.992 & 0.940 & 0.896 & 0.925 & 0.355 & 0.914 & 0.125 \\
MY29 & H6 & 0.990 & 0.921 & 0.799 & 0.896 & 0.321 & 0.886 & 0.115 \\
MY30 & H1 & 0.998 & 0.978 & 0.959 & 0.972 & 0.722 & 0.986 & 0.734 \\
MY30 & H3 & 0.994 & 0.977 & 0.873 & 0.953 & 0.521 & 0.972 & 0.592 \\
MY30 & H6 & 0.992 & 0.964 & 0.783 & 0.934 & 0.505 & 0.949 & 0.515 \\
MY31 & H1 & 0.995 & 0.813 & 0.959 & 0.953 & 0.565 & 0.907 & 0.065 \\
MY31 & H3 & 0.991 & 0.944 & 0.884 & 0.942 & 0.380 & 0.925 & 0.158 \\
MY31 & H6 & 0.988 & 0.928 & 0.809 & 0.921 & 0.335 & 0.904 & 0.141 \\
MY32 & H1 & 0.995 & 0.817 & 0.951 & 0.948 & 0.544 & 0.907 & 0.049 \\
MY32 & H3 & 0.991 & 0.942 & 0.864 & 0.931 & 0.353 & 0.921 & 0.128 \\
MY32 & H6 & 0.988 & 0.924 & 0.784 & 0.902 & 0.308 & 0.894 & 0.108 \\
MY34 & H1 & 0.993 & 0.288 & 0.908 & 0.497 & -0.093 & 0.859 & 0.088 \\
MY34 & H3 & 0.983 & 0.428 & 0.812 & 0.612 & -1.676 & 0.869 & -0.538 \\
MY34 & H6 & 0.984 & 0.421 & 0.691 & 0.611 & -0.847 & 0.809 & -0.079 \\
\hline
\end{tabular}
\end{table*}

\paragraph{Global Dust Storm Regime Behaviour in MY34}

\begin{figure*}[ht]
\centering
\includegraphics[width=0.95\textwidth]{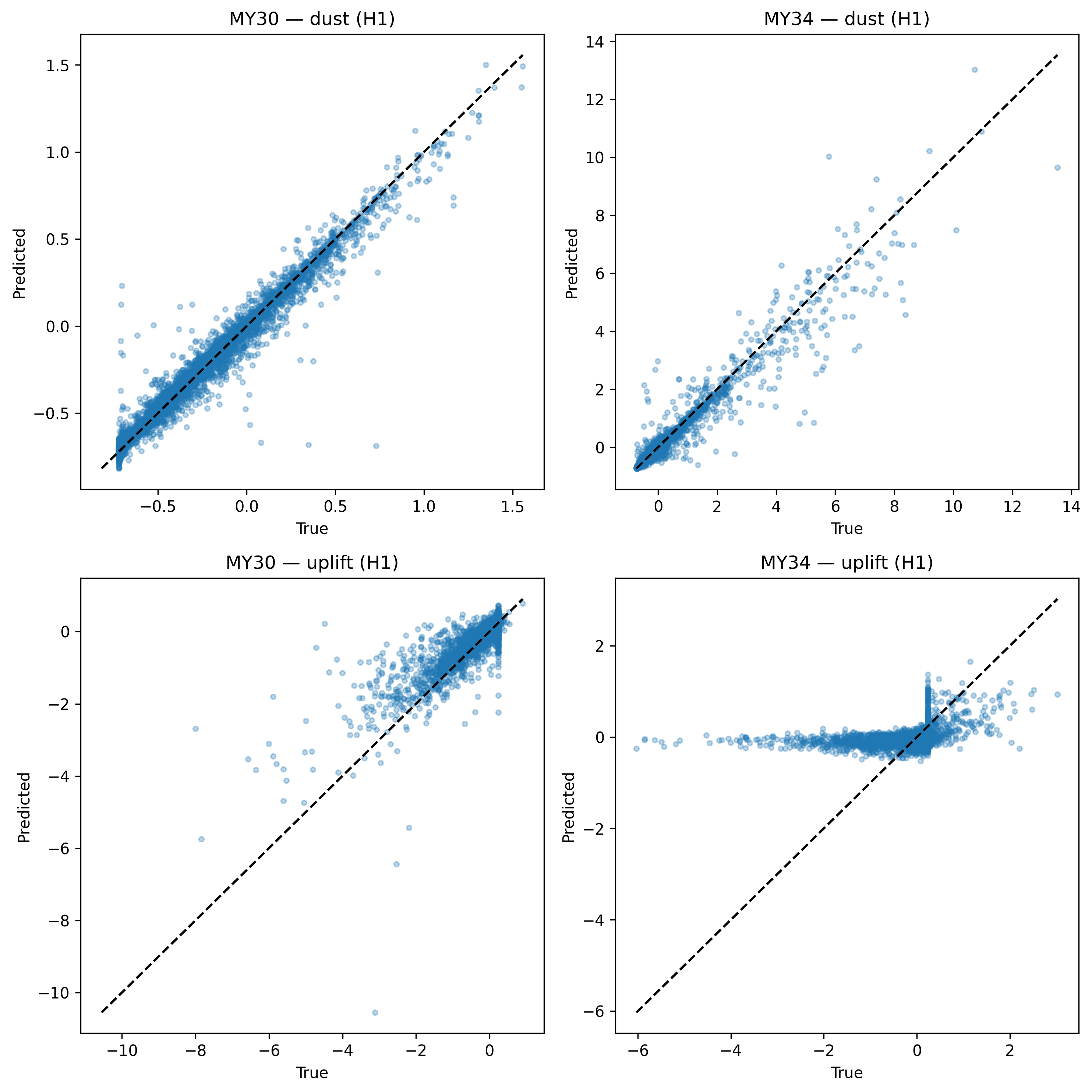}
\caption{Truth-versus-prediction scatter plots for MY30 and MY34 at H1, comparing dust-column and uplift forecasts. Dust-column predictions remain closely aligned with the one-to-one reference line, while uplift predictions show weaker agreement, particularly under the MY34 global dust storm regime.}
\label{fig:my30_my34_dust_uplift_h1_scatter}
\end{figure*}

Figure~\ref{fig:my30_my34_dust_uplift_h1_scatter} provides a visual comparison of H1 dust-column and uplift predictions for MY30 and MY34. The dust-column predictions remain closely aligned with the one-to-one line in both years, supporting the strong dust \(R^2\) values reported in Table~\ref{tab:r2_horizon_variable}. In contrast, uplift predictions are much less stable, especially in MY34, where the model tends to compress the predicted uplift range. This supports the interpretation that uplift is a more difficult derived diagnostic target than dust column.

The MY34 results show clear regime-dependent behaviour. In Table~\ref{tab:overall_generalisation}, MY34 has the weakest overall performance, with \(R^2 = 0.363\). This contrasts with the regular-year range of 0.726--0.851. However, Table~\ref{tab:r2_horizon_variable} shows that dust-column prediction remains relatively strong in MY34, with dust \(R^2\) values of 0.908, 0.812, and 0.691 for H1, H3, and H6. As defined in Section~\ref{sec:graph}, H1, H3, and H6 correspond to target windows shifted by 1, 3, and 6 graph chunks, respectively.

This indicates that the MY34 degradation is not caused by a complete failure of dust-column prediction. Instead, the weaker overall score is driven by poorer performance in other variables, especially meridional wind \(v\), uplift, and surface temperature. The MY34 case therefore acts as a useful stress test: the model retains meaningful dust forecasting skill but struggles with the broader multivariate atmospheric regime.

\paragraph{Auxiliary Dust-Regime Classification}

Table~\ref{tab:cas_classification} reports the performance of the auxiliary dust-regime classification head. This head is trained using a weak binary dust-active label derived from the dust-threshold flag created during graph construction. It is therefore not an independently validated dust storm detector. Instead, it is used as an auxiliary regime signal that helps the model identify locally dust-active atmospheric states and condition the adaptive routing mechanism.

The auxiliary classifier performs strongly across the regular evaluation years, with accuracy values between 0.995 and 1.000 and AUC values of 1.000. MY34 behaves differently. Accuracy falls to 0.627, while precision, recall, and F1 collapse to approximately zero, and AUC decreases to 0.200. This indicates that the weak threshold-derived dust-regime signal does not transfer cleanly to the MY34 global dust storm evaluation case.

This result should be interpreted alongside the dust-column regression results. The model retains strong MY34 dust-column forecasting skill, but the auxiliary classification head fails to identify the MY34 dust-active regime using the same weak binary target. This pattern is consistent with a regime-transfer problem in the auxiliary classifier, rather than a complete failure of dust-column forecasting.

\begin{table}[ht]
\centering
\caption{Auxiliary dust-regime classification performance}
\label{tab:cas_classification} 
\begin{tabular}{cccccc}
\hline
Year & Acc & Prec & Recall & F1 & AUC \\
\hline
MY29 & 0.998 & 0.994 & 0.999 & 0.996 & 1.000 \\
MY30 & 1.000 & 1.000 & 1.000 & 1.000 & 1.000 \\
MY31 & 0.995 & 0.986 & 0.998 & 0.992 & 1.000 \\
MY32 & 0.997 & 0.992 & 0.998 & 0.995 & 1.000 \\
MY34 & 0.627 & 0.001 & 0.000 & 0.000 & 0.200 \\
\hline
\end{tabular}
\end{table}

\paragraph{Regional Forecast Heterogeneity}

\begin{figure*}[ht]
\centering
\includegraphics[width=0.95\textwidth]{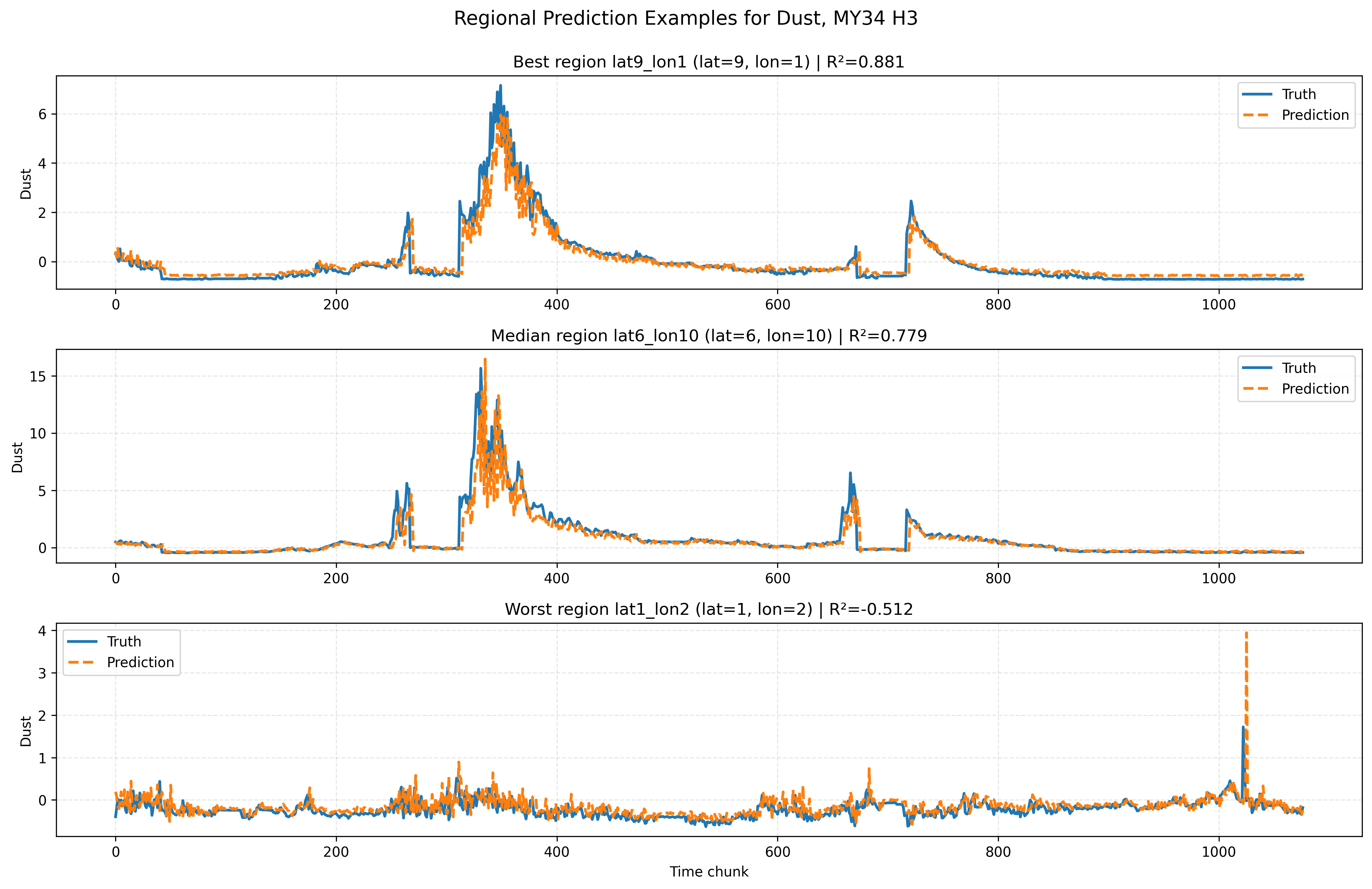}
\caption{Regional prediction examples for MY34 dust-column forecasting at H3. The panels show representative best-case, median-case, and worst-case spatial regions. The best and median regions demonstrate that the model can track the main dust evolution in some locations, while the worst-case region illustrates a regional failure mode with a false positive late spike.}
\label{fig:regional_examples_my34_dust_h3}
\end{figure*}

Figure~\ref{fig:regional_examples_my34_dust_h3} shows example regional dust-column forecasts for MY34 at H3. The selected regions illustrate best-case, median-case, and worst-case regional behaviour. The best and median regions show that the model captures the main dust-loading evolution and decay structure, although peak amplitudes are sometimes underestimated. The worst-case region shows a different failure mode, where the model produces a late false positive spike that is not present in the target series. These examples show that MY34 forecast skill is spatially heterogeneous: the model can track dust evolution well in some regions, while other regions remain difficult under the global dust storm regime.

\paragraph{Inference-Time Component Sensitivity}

Inference-time component sensitivity was used to examine which information pathways the trained model depends on most strongly. This analysis was performed by evaluating the trained model after masking or removing selected inputs or graph relation groups at inference time. It should therefore be interpreted as a diagnostic sensitivity test rather than a fully retrained ablation study.

Table~\ref{tab:component_sensitivity} shows that the raw temporal sequence pathway is the dominant contributor to performance. Removing the temporal sequence features causes mean \(R^2\) to fall from 0.681 to \(-0.016\), corresponding to a \(\Delta R^2\) of \(-0.698\). Dust H3 performance also collapses, with dust H3 \(R^2\) falling from 0.866 to \(-0.193\). This indicates that recent temporal atmospheric history is the most important information source for the trained model.

Graph structure also contributes positively. Removing all graph edges reduces mean \(R^2\) from 0.681 to 0.618, giving a \(\Delta R^2\) of \(-0.063\). Removing temporal graph edges causes a smaller overall degradation of \(-0.036\), while removing similarity edges causes only a small degradation of \(-0.005\). These results suggest that graph propagation improves performance, but its contribution is secondary to the raw temporal sequence representation.

The engineered/static feature branch also contributes to overall performance, with removal reducing mean \(R^2\) by \(-0.026\). In contrast, removing the vertical column tensor has very little effect on the aggregate metrics, with mean \(R^2\) changing from 0.681 to 0.680. This result indicates that, despite the additional complexity of the vertical encoder, the current trained model makes only limited use of the full vertical tensor under this sensitivity test.

Several aspects of the current representation may contribute to this weak marginal effect. First, compact vertical summary statistics are already included in the engineered/static feature branch, meaning that some vertical information is available to the model even when the full vertical tensor is removed. Second, the full vertical profile is taken only from the final time step of each node window, whereas the temporal branch receives an eight-step history. Third, variables without an intrinsic vertical dimension, including surface pressure and dust column, are broadcast across levels to maintain a consistent tensor shape and therefore do not contribute additional vertical variation. The vertical pathway may consequently contain information that is partly redundant with the temporal and engineered branches.

This finding does not imply that vertical atmospheric structure is physically unimportant. It instead shows that the present implementation does not extract a strong additional predictive benefit from the vertical encoder for the forecasting targets considered here. Because the analysis is an inference-time sensitivity test rather than a fully retrained ablation, future work should directly compare retrained models with and without the vertical pathway and investigate alternative vertical representations, including temporally evolving vertical profiles and simpler vertical encoders, to determine whether the additional architectural complexity is warranted.

Finally, removing spatial graph edges slightly increases mean overall \(R^2\) from 0.681 to 0.688, while reducing dust H3 \(R^2\) from 0.866 to 0.861. This mixed result suggests that local spatial propagation may slightly oversmooth some variables while still contributing modestly to dust-column forecasting. Overall, the sensitivity analysis shows that the model is primarily driven by temporal sequence information, with graph structure and engineered features providing secondary gains.

\begin{table*}[ht]
\centering
\caption{Inference-time component sensitivity analysis. Negative \(\Delta R^2\) values indicate performance degradation relative to the full model.}
\label{tab:component_sensitivity}
\begin{tabular}{lrrrr}
\hline
Variant & Mean \(R^2\) & \(\Delta R^2\) & Dust H3 \(R^2\) & Dust H3 \(\Delta R^2\) \\
\hline
No temporal sequence & -0.016 & -0.698 & -0.193 & -1.059 \\
No graph edges & 0.618 & -0.063 & 0.854 & -0.012 \\
No temporal graph edges & 0.645 & -0.036 & 0.871 & 0.005 \\
No engineered/static features & 0.656 & -0.026 & 0.855 & -0.011 \\
No similarity edges & 0.676 & -0.005 & 0.868 & 0.002 \\
No vertical column & 0.680 & -0.001 & 0.866 & 0.000 \\
Full model & 0.681 & 0.000 & 0.866 & 0.000 \\
No spatial graph edges & 0.688 & 0.006 & 0.861 & -0.005 \\
\hline
\end{tabular}
\end{table*}

\paragraph{Dust-Column Baseline Comparison Against Classical Models}

Table~\ref{tab:classical_dust_baseline} reports the dust-column forecasting comparison against the classical baselines. The GNN achieves the best \(R^2\) in 13 of the 15 year--horizon comparisons. The only cases where the GNN is not the best model are MY30 H1 and MY30 H3, where the difference is very small. In MY30 H1, Random Forest achieves \(R^2 = 0.960\), compared with \(R^2 = 0.959\) for the GNN. In MY30 H3, Linear Regression achieves \(R^2 = 0.881\), compared with \(R^2 = 0.873\) for the GNN.

Across the full dust-column comparison, the mean GNN \(R^2\) is 0.863. This compares with a mean Random Forest \(R^2\) of 0.618 and a mean Linear Regression \(R^2\) of 0.853 when excluding the unstable MY34 out-of-distribution rows. The mean gain over the strongest stable classical baseline is 0.042.

The MY34 results are particularly important because this year acts as the global dust storm stress-test case. Linear Regression produced unstable MY34 extrapolations and is therefore omitted from those rows. Against the stable Random Forest baseline, the GNN improves dust-column \(R^2\) at all MY34 horizons. The gain is modest at H1 \((+0.012)\), larger at H3 \((+0.093)\), and strongest at H6 \((+0.226)\). This indicates that the GNN retains stronger dust-column forecasting skill than the classical baselines under the global dust storm regime, especially at longer forecast horizons.

\begin{table*}[ht]
\centering
\caption{Dust-column forecasting baseline comparison measured using \(R^2\). Gain is computed relative to the strongest stable classical baseline for each year--horizon pair. Linear Regression produced unstable out-of-distribution extrapolations under MY34 and is omitted from those rows.}
\label{tab:classical_dust_baseline}
\begin{tabular}{llrrrr}
\hline
Martian Year & Horizon & Linear Regression & Random Forest & GNN & GNN Gain \\
\hline
MY29 & H1 & 0.920 & 0.532 & 0.973 & 0.053 \\
MY29 & H3 & 0.857 & 0.616 & 0.896 & 0.039 \\
MY29 & H6 & 0.772 & 0.556 & 0.799 & 0.027 \\
MY30 & H1 & 0.952 & 0.960 & 0.959 & -0.001 \\
MY30 & H3 & 0.881 & 0.880 & 0.873 & -0.008 \\
MY30 & H6 & 0.776 & 0.751 & 0.783 & 0.007 \\
MY31 & H1 & 0.908 & 0.413 & 0.959 & 0.051 \\
MY31 & H3 & 0.855 & 0.541 & 0.884 & 0.029 \\
MY31 & H6 & 0.793 & 0.516 & 0.809 & 0.016 \\
MY32 & H1 & 0.902 & 0.384 & 0.951 & 0.049 \\
MY32 & H3 & 0.841 & 0.525 & 0.864 & 0.023 \\
MY32 & H6 & 0.774 & 0.510 & 0.784 & 0.010 \\
MY34 & H1 & -- & 0.896 & 0.908 & 0.012 \\
MY34 & H3 & -- & 0.719 & 0.812 & 0.093 \\
MY34 & H6 & -- & 0.465 & 0.691 & 0.226 \\
\hline
\end{tabular}
\end{table*}

\paragraph{Dust-Column Baseline Comparison Against Deep Temporal Models}

Table~\ref{tab:deep_temporal_dust_baseline} reports the dust-column forecasting comparison against the deep temporal baselines. These baselines provide a stronger test than the classical models because LSTM, GRU, and TCN models are designed to learn temporal structure directly. The GNN achieves the best dust-column \(R^2\) in 14 of the 15 year--horizon comparisons.

The only case where the GNN is not the best model is MY34 H1. In that case, GRU achieves \(R^2 = 0.911\), while the GNN achieves \(R^2 = 0.908\). This difference is small. At the longer MY34 horizons, the GNN performs best, with \(R^2 = 0.812\) at H3 and \(R^2 = 0.691\) at H6.

Across all dust-column comparisons, the mean GNN dust \(R^2\) is 0.863. This compares with 0.800 for LSTM, 0.831 for GRU, and 0.769 for TCN. These results show that the graph-based model provides stronger dust-column forecasting skill than the temporal-only baselines, especially at medium and longer forecast horizons.

It is important to note that this table focuses specifically on dust-column forecasting. The overall multivariate comparison shows that temporal baselines perform better than the GNN on aggregate MY34 metrics. However, for the central dust-column forecasting task, the GNN remains the strongest model in nearly all year--horizon cases.

\begin{table*}[ht]
\centering
\caption{Dust-column forecasting comparison against deep temporal baselines measured using \(R^2\).}
\label{tab:deep_temporal_dust_baseline}
\begin{tabular}{llrrrr}
\hline
Martian Year & Horizon & LSTM & GRU & TCN & GNN \\
\hline
MY29 & H1 & 0.943 & 0.947 & 0.938 & 0.973 \\
MY29 & H3 & 0.853 & 0.872 & 0.845 & 0.896 \\
MY29 & H6 & 0.725 & 0.766 & 0.684 & 0.799 \\
MY30 & H1 & 0.933 & 0.946 & 0.933 & 0.959 \\
MY30 & H3 & 0.849 & 0.868 & 0.820 & 0.873 \\
MY30 & H6 & 0.681 & 0.707 & 0.558 & 0.783 \\
MY31 & H1 & 0.920 & 0.930 & 0.911 & 0.959 \\
MY31 & H3 & 0.829 & 0.858 & 0.817 & 0.884 \\
MY31 & H6 & 0.691 & 0.755 & 0.635 & 0.809 \\
MY32 & H1 & 0.918 & 0.927 & 0.909 & 0.951 \\
MY32 & H3 & 0.824 & 0.850 & 0.812 & 0.864 \\
MY32 & H6 & 0.695 & 0.756 & 0.641 & 0.784 \\
MY34 & H1 & 0.889 & 0.911 & 0.891 & 0.908 \\
MY34 & H3 & 0.746 & 0.790 & 0.703 & 0.812 \\
MY34 & H6 & 0.506 & 0.588 & 0.433 & 0.691 \\
\hline
\end{tabular}
\end{table*}

\subsection{Cross-Planet Framework Transferability}

To demonstrate that the proposed graph construction and forecasting framework has transferability outside of the OpenMARS dataset the architectural pipeline was also demonstrated on two additional datasets. This is a supplementary demonstration rather than a main result. The aim was to test whether the same data engineering graph based approach could be reused with a different atmospheric grid, variable set.

For Venus, the graph construction used the same general representation as the Mars experiment, using the Venus Planetary Climate Model output dataset for age of air from the surface~\cite{cohen2024venusAgeAirSurface}: local spatiotemporal nodes, raw temporal sequence features, vertical column tensors, spatial graph edges, temporal graph edges, and similarity edges. The model architecture was also kept, with temporal encoding, vertical encoding, relational graph propagation, adaptive routing, and multi-horizon forecast heads.

\begin{figure*}[ht]
\centering
\includegraphics[width=0.85\textwidth]{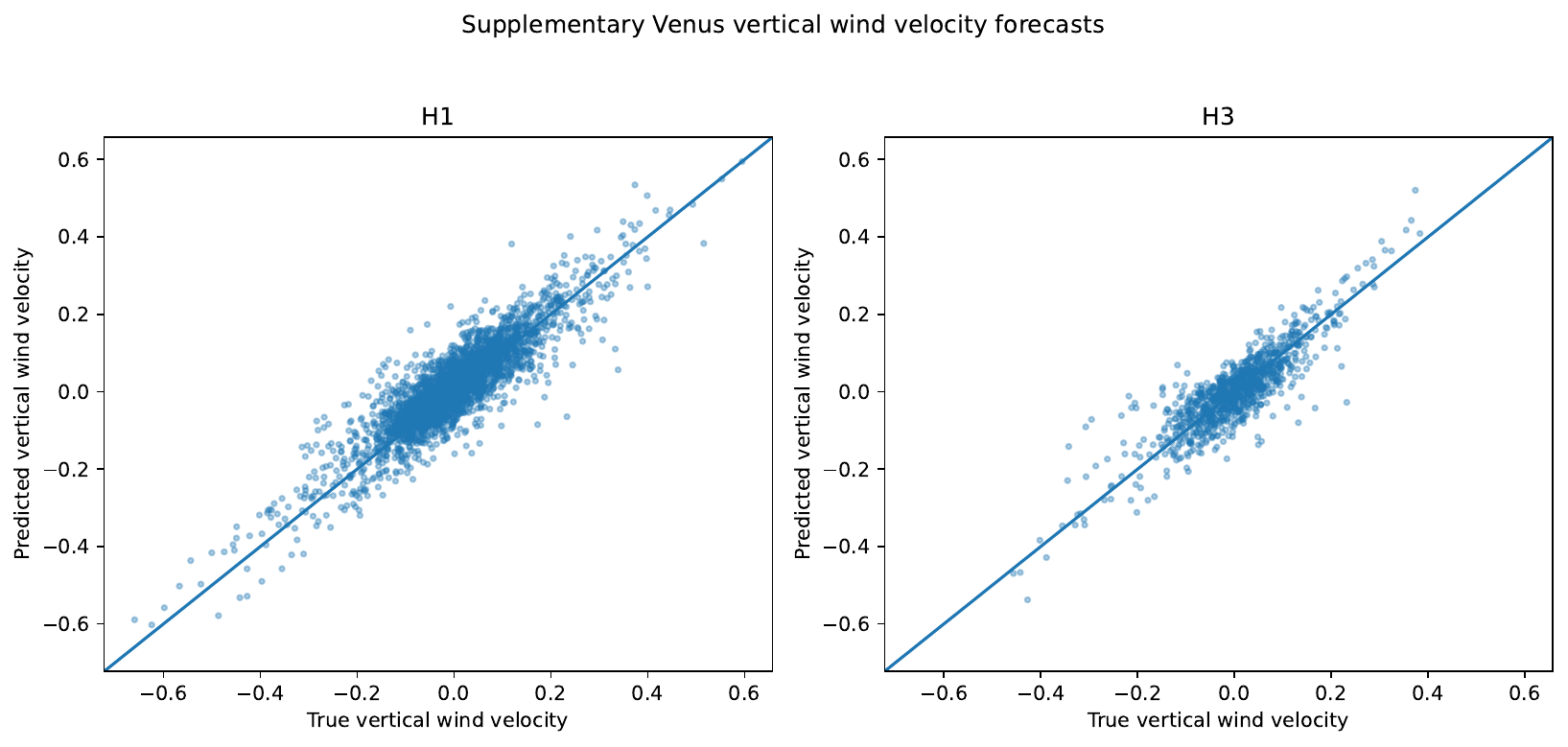}
\caption{Supplementary Venus truth-versus-prediction forecasts for vertical wind velocity at H1 and H3. The plotted field corresponds to \texttt{vitw} in the Venus dataset. This figure is included as an ability demonstration showing that the graph construction and multi-horizon forecasting pipeline can be used on a non-Martian planetary atmospheric dataset.}
\label{fig:supp_venus_vertical_wind_h1_h3}
\end{figure*}

Figure~\ref{fig:supp_venus_vertical_wind_h1_h3} shows the supplementary Venus truth-versus-prediction result for vertical wind velocity at H1 and H3. The H1 forecast achieved \(R^2 = 0.774\), showing that the framework can produce meaningful short-horizon forecasts when used on a non-Martian planetary atmospheric dataset. This result is not a claim of a complete Venus forecasting model, but it shows the broader transferability of the graph and data engineering approach to atmospheric weather forecasting.

Future supplementary experiments will extend this cross-planet demonstration to additional atmospheric datasets, including the Titan polar-cloud simulation dataset~\cite{deBatz2025TitanPolarCloudsDataset}, where the same graph-construction principles can be tested under a different atmospheric regime, variable set, and dynamical structure.

\begin{figure}[!htbp]
\centering
\includegraphics[width=0.95\textwidth]{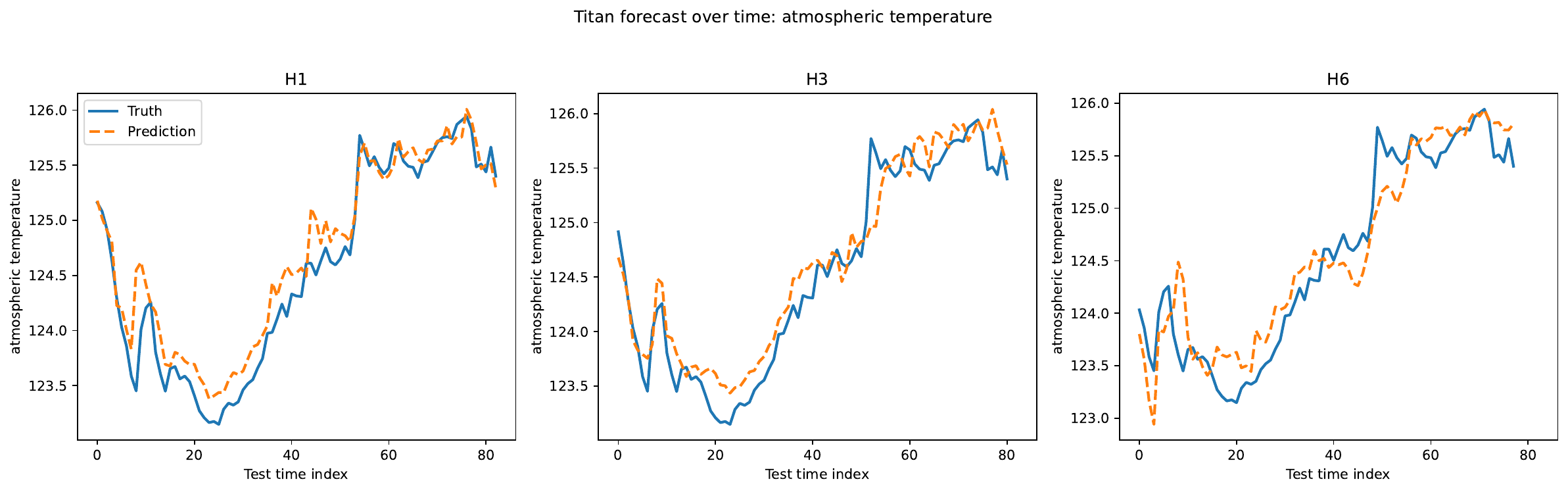}
\caption{Supplementary Titan forecast demonstration showing truth and predicted temporal evolution for atmospheric temperature across the H1, H3, and H6 forecast heads. Each panel has the spatial mean over the Titan test graph at each test time step. This figure is included as a cross-planet portability demonstration of the graph-construction and multi-horizon forecasting framework rather than as a full Titan atmospheric forecasting study.}
\label{fig:supp_titan_temperature_forecast}
\end{figure}
Figure~\ref{fig:supp_titan_temperature_forecast} shows an additional Titan portability demonstration using atmospheric temperature. The model tracks the spatially averaged temporal evolution across H1, H3, and H6, showing that the same graph-construction and multi-horizon forecasting framework can be used on another planetary atmospheric dataset.

\subsection{Summary of Findings}

The experiments show that MaGMA provides a useful graph-based representation for Martian atmospheric forecasting. Across the regular unseen Martian years MY29--MY32, the model achieves stable overall performance, with $R^2$ values between 0.726 and 0.851. The strongest results are observed for variables with clear temporal and large-scale atmospheric structure, including surface pressure, dust column, atmospheric temperature, and zonal wind.

The MY34 global dust storm year provides a more demanding test. Overall multivariate performance drops substantially, showing that transfer from one dust-storm year to another extreme regime remains difficult. However, the degradation is not uniform. Dust-column prediction remains strong at H1 and H3, while meridional wind, uplift, and surface temperature are harder to forecast under MY34. This indicates that the model retains useful dust-loading skill during the global dust storm regime, while still struggling with the broader coupled atmospheric response.

The baseline comparisons further support the graph-based approach. For dust-column forecasting, MaGMA outperforms the classical baselines in most year--horizon comparisons and outperforms the deep temporal baselines in nearly all comparisons. The advantage is clearest at medium and longer horizons, especially under MY34, where the graph-based model retains stronger dust-column skill than the temporal-only baselines.

The sensitivity analysis also clarifies how the model uses the representation. Recent temporal history is the dominant information source: removing the raw temporal sequence causes the largest performance collapse. Graph structure provides a smaller but still useful contribution, suggesting that MaGMA does not replace temporal forecasting but strengthens it with spatial, temporal, and dynamically similar atmospheric context. The vertical pathway contributes less than expected in the current implementation, indicating that future work should investigate stronger vertical encoders and more targeted use of column information, especially during dust-active periods.

Finally, the Venus and Titan demonstrations show that the graph-construction pipeline is not tied only to OpenMARS. These experiments are supplementary rather than complete forecasting studies, but they show that the same representation principles can be reused with atmospheric datasets that differ in variable structure, physical regime, and planetary context.

\section{Conclusion}
\label{sec:conclusion}
This paper introduced MaGMA, a graph-based data engineering and forecasting framework for planetary atmospheric data. The central contribution is the transformation of multidimensional reanalysis fields into graph-structured learning objects that represent local atmospheric patches, recent histories, vertical information, spatial neighbourhoods, temporal links, and dynamically similar atmospheric states together. This changes the forecasting problem from isolated sequence prediction to structured atmospheric learning.

The results show that this representation is effective for multi-horizon Martian atmospheric forecasting. MaGMA performs consistently across regular unseen Martian years and provides strong dust-column forecasting, including during the MY34 global dust storm year at shorter horizons. At the same time, the MY34 results show that rare extreme regimes remain challenging, especially for variables such as meridional wind, uplift, and surface temperature. The framework therefore supports both forecasting and diagnosis: it identifies where graph-based learning improves prediction and where the coupled atmospheric response remains difficult to model.

The study also investigates how that representation design can be reused for machine learning in planetary science. The graph does not replace temporal information; rather, it augments recent atmospheric history with relational context across space, time, and atmospheric state similarity. The supplementary Venus and Titan demonstrations further indicate that the same graph-construction principles can be reused beyond Mars.

Future work should focus on improving generalisation under rare dust-storm regimes, making better use of vertical atmospheric structure, and extending the framework to additional planetary datasets and forecasting tasks. A further priority is benchmarking MaGMA against modern spatiotemporal atmospheric forecasting architectures, including graph-based global forecasting models, neural-operator approaches, and transformer-based weather models. The LSTM, GRU, and TCN baselines used in the present study were selected to test the contribution of graph structure beyond temporal-only forecasting, but they do not constitute an exhaustive comparison with current large-scale spatiotemporal forecasting systems. Such comparisons would require adapting these architectures to the OpenMARS grid, variable set, forecast horizons, and available training-data regime. More broadly, MaGMA shows how graph-based data engineering can provide a reusable and diagnostically useful bridge between scientific atmospheric data products and machine learning models.

\section{Data Availability}

The OpenMARS continuous MY28--35 reanalysis product used as the source dataset in this study is publicly available from The Open University repository at~\url{https://doi.org/10.21954/ou.rd.24573205}. The supplementary Venus dataset is publicly available at~\url{https://doi.org/10.21954/ou.rd.26062744}. The supplementary Titan dataset is publicly available at~\url{https://doi.org/10.14768/85b9e037-57a2-46b5-945b-23a02e32aca2}.

The code used to construct the graph datasets, train the MaGMA model, run the forecasting experiments, and reproduce the main workflow is available at~\url{https://github.com/gmyler/MaGMA---Martian-Graph-based-Multi-horizon-Atmospheric-Forecasting.git}.

\section*{Declaration on Generative AI}
The authors used OpenAI ChatGPT to support language polishing, clarity improvement, LaTeX formatting, code debugging, and workflow assistance. The authors reviewed all content and take full responsibility for the final manuscript.

\section*{Funding}
 This work was supported by the Engineering and Physical Sciences Research Council [grant numbers EP/V026747/1, EP/R013144/1].

\section*{Competing Interests}

The authors declare that they have no competing interests.

\bibliography{sn-bibliography,bibliography}

\end{document}